%% file: main.tex
\documentclass[twocolumn,prx,aps,superscriptaddress,showpacs]{revtex4-2}
\usepackage{CJK}
\usepackage{lineno}
\usepackage{graphicx}
\usepackage{amsmath}
\usepackage{bbold}
\usepackage{amssymb}
\usepackage[english]{babel}
\usepackage{color}
\usepackage[version=4]{mhchem}
\usepackage[hidelinks]{hyperref}
\usepackage{dsfont}
\usepackage{placeins}
\usepackage{mathtools}
\usepackage[normalem]{ulem}
\usepackage{lipsum}
\usepackage{array}
\usepackage{comment}
\usepackage{hyperref}
\usepackage{physics}
\hypersetup{colorlinks=true,citecolor=blue,linkcolor=blue,urlcolor=blue}

\usepackage{soul}
\newcommand{\Caltech}{California Institute of Technology, Pasadena, CA 91125, USA}

\newcommand{\highRabi}{$7.7$~MHz}

\newcommand{\highFidBBCorrected}{$0.9971(5)$}
\newcommand{\ground}{$^1\text{S}_0$}
\newcommand{\clock}{$^3\text{P}_0$}
\newcommand{\rstate}{$^3\text{P}_1$}
\newcommand{\mstate}{$^3\text{P}_2$}

\usepackage{caption}

\DeclareCaptionLabelSeparator{bar}{ \textbf{\textbar}~}
\usepackage{enumitem}
\setlist{nolistsep}
\begin{document}
\title{Benchmarking and fidelity response theory of high-fidelity Rydberg entangling gates}
\author{Richard Bing-Shiun Tsai}
\thanks{These authors contributed equally to this work.}
\author{Xiangkai Sun}
\thanks{These authors contributed equally to this work.}
\author{Adam L. Shaw}
\affiliation{\Caltech}
\author{Ran Finkelstein}\email{rfinkel@caltech.edu}
\author{Manuel Endres}\email{mendres@caltech.edu}
\affiliation{\Caltech}

\begin{abstract}
The fidelity of entangling operations is a key figure of merit in quantum information processing, especially in the context of quantum error correction.
High-fidelity entangling gates in neutral atoms have seen remarkable advancement recently.
A full understanding of error sources and their respective contributions to gate infidelity will enable the prediction of fundamental limits on quantum gates in neutral atom platforms with realistic experimental constraints.
In this work, we implement the time-optimal Rydberg CZ gate, design a circuit to benchmark its fidelity, and achieve a fidelity, averaged over symmetric input states, of \highFidBBCorrected, downward-corrected for leakage error, which together with our recent work forms a new state-of-the-art for neutral atoms.
The remaining infidelity is explained by an \textit{ab initio} error model, consistent with our experimental results over a range of gate speeds, with varying contributions from different error sources.
Further, we develop a fidelity response theory to efficiently predict infidelity from laser noise with non-trivial power spectral densities and derive scaling laws of infidelity with gate speed.
Besides its capability of predicting gate fidelity, we also utilize the fidelity response theory to compare and optimize gate protocols, to learn laser frequency noise, and to study the noise response for quantum simulation tasks.
Finally, we predict that a CZ gate fidelity of ${\gtrsim} 0.999$ is feasible with realistic experimental upgrades.
\end{abstract}

\maketitle
\section{Introduction}
A central challenge in advancing quantum technologies is lowering error rates below the threshold for quantum error correction~\cite{Fowler2012}, so as to utilize the potential of large-scale processors, which are currently limited largely by entangling gate fidelities. Two-qubit gates mediated by Rydberg interactions~\cite{Jaksch2000,Saffman2010,Isenhower2010,Wilk2010} have enabled high-fidelity quantum logic to be realized in neutral atom platforms. In particular, recent improvements in gate protocols~\cite{Jandura2022, Pagano2022} have resulted in the realization of entangling gate fidelities $>99.5\%$~\cite{Evered2023, Finkelstein2024} and high-fidelity operation across different atomic species~\cite{Evered2023, Finkelstein2024, Peper2024, Cao2024} --- bringing neutral atoms closer to other leading quantum computing platforms, such as trapped ions~\cite{Moses2023,Loschnauer2024} and superconducting qubits~\cite{Acharya2023,Zhang2022}. A further increase in gate fidelity could contribute to an exponential suppression of logical error rates~\cite{Fowler2012}. It is thus important to accurately benchmark entangling gate fidelities, to model mechanisms that plague neutral atom processors, and to predict how to overcome these.

\begin{figure*}[ht!]
    \centering
    \includegraphics[width=\textwidth]{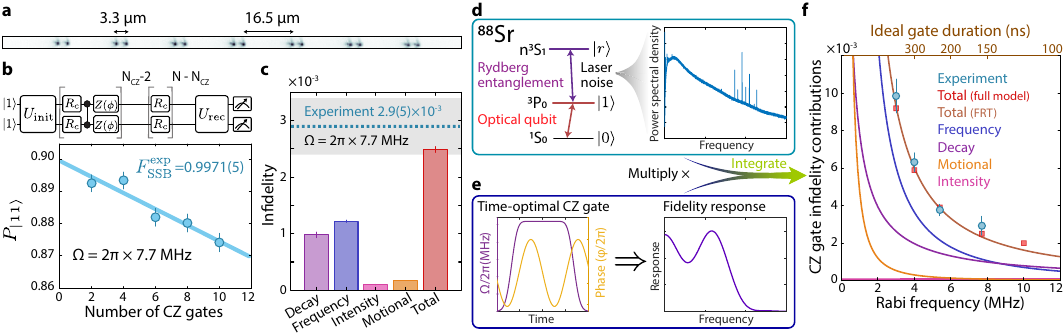}
    \caption{\textbf{Summary of results.}
    \textbf{(a)} Array configuration of tweezer-trapped strontium atoms, depicted by an average atomic fluorescence image, for parallel CZ gate implementation.
    \textbf{(b)} Circuit for the symmetric stabilizer benchmarking (SSB) sequence and experimental results for a CZ gate with a Rydberg Rabi frequency of \highRabi. More details on the circuit implementation are given in Fig.~\ref{SI_SSB_seq}. By fitting an exponential decay to $P_{\ket{11}}$ as a function of $N_\mathrm{CZ}$, the total number of CZ gates applied (Appendix~\ref{Appendix:Setup}), we infer a CZ fidelity of \highFidBBCorrected, averaged over symmetric Haar random states and corrected for false contribution from leakage (Appendix~\ref{Appendix:BrightState}).
    \textbf{(c)} From a full \textit{ab initio} error model simulation at a Rydberg Rabi frequency of \highRabi, we predict contributions of dominant error sources, which include Rydberg decay, the frequency and intensity noise of the Rydberg laser, and atomic motion (see Appendix~\ref{Appendix:Setup} for details on the effects of tweezer traps). The total predicted infidelity is within error bars of the experimental result (dashed line and shaded area).
    \textbf{(d)} The relevant electronic states of strontium-88 atoms. The Rydberg laser intensity and frequency noise, perturbing the execution of an ideal CZ gate, is characterized by a power spectral density (PSD).
    \textbf{(e)} From the pulse shape of the time-optimal CZ gate and the ideal trajectory in the corresponding Hilbert space, we derive a fidelity response to the specific noise source PSD.
    \textbf{(f)} Combining the fidelity response function and the noise PSD, we compute the infidelity stemming separately from various error sources as a function of Rydberg Rabi frequency during an ideal time-optimal CZ gate, including laser frequency noise (blue), Rydberg decay (purple), atomic motion (orange), and laser intensity noise (pink). Summing them up, we obtain a CZ gate infidelity prediction over symmetric Haar random states (brown) from the fidelity response theory (FRT). Full error model simulation results (red squares) are within error bars of experimental results (light blue circles). A slight deviation between the full error model and FRT at the highest Rabi frequency stems mainly from acousto-optical modulators (AOM) rise and fall time effects (see text). Ideal gate duration is indicated, assuming zero AOM rise/fall time.
    }
    \vspace{-0.3cm}
    \label{fig:overview}
\end{figure*}

In this work, we benchmark the fidelity of the controlled-Z (CZ) gate in an array of tweezer-trapped strontium-88 atoms~\cite{Cooper2018,Norcia2018, Madjarov2019,Norcia2019,Madjarov2020} (Fig.~\ref{fig:overview}a) with a symmetric stabilizer benchmarking (SSB) sequence, designed to benchmark the gate fidelity averaged over the two-qubit symmetric subspace. This specific design is motivated by recent benchmarking sequences realized for neutral atoms~\cite{Evered2023,Ma2023} and the need to individually characterize the native entangling CZ gate, mediated by Rydberg interaction~\cite{Browaeys2020}. From the sequence, we obtain a CZ gate fidelity of \highFidBBCorrected, downward-corrected for leakage error (Appendix~\ref{Appendix:BrightState}), at the highest achievable Rydberg Rabi frequency of \highRabi\ on our current setup (Fig.~\ref{fig:overview}b, c). We study in depth an \textit{ab initio} error model and find good agreement between our measured results and numerical predictions across a wide range of applied Rabi frequencies. 

Moreover, to gain a more intuitive, analytical understanding of how error sources contribute to the total infidelity, we develop a fidelity response theory (FRT) to analyze the effect of non-Markovian noise with a non-trivial power spectral density (PSD), as is particularly relevant for laser noise affecting Rydberg atom array systems~\cite{deLeseleuc2018, Jiang2023,Shaw2024C} (Fig.~\ref{fig:overview}d-f). Summing up analytical FRT predictions of infidelity contributions from each error source, we find good agreement between predicted CZ gate infidelity and the full numerical \textit{ab initio} error model prediction. This suggests that FRT is a reliable toolbox to study Rydberg-mediated entangling gates.

FRT further provides a shortcut to understanding the noise response in a much wider range of scenarios. We demonstrate this explicitly by predicting performance for different gate protocols (including addressing different Rydberg states, optimizing pulse shapes, and generalizing to two-photon transitions in alkali atoms~\cite{Isenhower2010,Levine2019}), probing laser frequency noise from atomic signals, and studying the response to laser noise in many-body systems.

Finally, guided by FRT and the good agreement between our experimental result and the \textit{ab initio} error model, we project a pathway for reaching ${\gtrsim} 0.999$ two-qubit gate fidelity in the near future with realistic experimental upgrades. 

The work is organized as follows:
In Section~\ref{Sec:benchmarking}, we introduce a new gate fidelity benchmarking method for neutral atom array experiments, requiring only global control. With this, we experimentally benchmark the time-optimal CZ gate~\cite{Jandura2022,Evered2023} fidelity at different Rabi frequencies. The benchmarking sequence is further accompanied by and justified with numerical studies using an \textit{ab initio} error model.
In Section~\ref{Sec:linear_response}, we develop FRT for predicting gate fidelities from noise power spectral densities.
In Section~\ref{Sec:application}, we describe several applications of FRT.
Finally, in Section~\ref{Sec:to999}, based on the toolbox developed in this work, we propose near-term experimental upgrades to reach ${\gtrsim}0.999$ CZ gate fidelity.

\section{Benchmarking high-fidelity entangling gates}\label{Sec:benchmarking}
\subsection*{Experimental implementation of entangling gates}
\begin{figure*}[ht!]
    \centering
    \includegraphics[width=\textwidth]{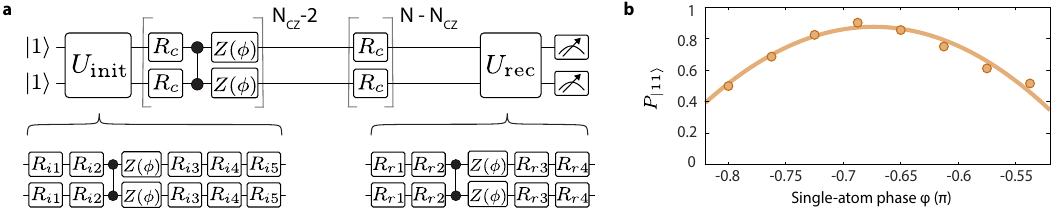}
    \caption{\textbf{Symmetric stabilizer benchmarking (SSB) sequence and single-atom phase calibration.}
    \textbf{(a)} An initialization unitary $\hat{U}_\mathrm{init}$ (decomposed into five $\pi/2$ pulses $R_i$ and one CZ gate) initializes each qubit pair to one of the twelve symmetric stabilizer states (SSS) (Table~\ref{tab:SSS}). Each subsequent rotation $R_C$ is randomly drawn from a set of four $\pi/2$ pulses with different phases. Among these rotations, the first $N_\mathrm{CZ}-2$ are interleaved with CZ gates. A pre-calculated recovery unitary $\hat{U}_\mathrm{rec}$ (decomposed into four $\pi/2$ pulses $R_r$ and one CZ gate) brings the state back to $\ket{11}$ in the absence of errors. The total number of applied CZ gates $N_\mathrm{CZ}$ (including the two CZ gates in $\hat{U}_\mathrm{init}$ and $\hat{U}_\mathrm{rec}$) is varied while $N$, the total number of randomly drawn $R_C$, is kept fixed. Note that after each Rydberg pulse, the qubits acquire a single-atom phase~\cite{Evered2023, Ma2023}. This single-atom phase is compensated by a virtual $Z(\phi)$ gate, imparted on the global clock laser phase after each Rydberg pulse. To emphasize this point, we show the global laser phase shift separately in the circuit, but this virtual $Z(\phi)$ gate is, in practice, a part of the CZ gate.
    \textbf{(b)} To calibrate this single-atom phase $\phi$ in the virtual $Z(\phi)$ gate after each Rydberg pulse, we perform the SSB sequence with $N_{\text{CZ}} = N = 10$ while scanning this parameter. The single-atom phase is then calibrated to maximize the return probability $P_{\ket{11}}$. 
    } 
    \vspace{-0.3cm}
    \label{SI_SSB_seq}
\end{figure*}

On our experiment, we trap individual strontium-88 atoms~\cite{Cooper2018,Norcia2018} and encode a long-lived qubit on an ultranarrow optical clock transition~\cite{Madjarov2019,Norcia2019} ($^1\text{S}_0$ $\leftrightarrow$ $^3\text{P}_0$) (Fig.~\ref{fig:overview}d and Appendix~\ref{Appendix:Setup}), addressed by a \textit{clock laser}. To generate entanglement between qubits, we use a \textit{Rydberg laser} to excite atoms from $^3\text{P}_0$ to the Rydberg state ($n^3\text{S}_1$, $n=61$)~\cite{Madjarov2020}. These highly excited states exhibit strong inter-atomic interactions. When two atoms are close enough, Rydberg interaction prevents the simultaneous Rydberg excitation of the two, a phenomenon commonly referred to as the Rydberg blockade~\cite{Browaeys2020}. This conditional excitation is the key ingredient for implementing a conditional two-qubit gate~\cite{Jaksch2000,Saffman2010}.

We implement a CZ gate~\cite{Graham2019,Levine2019} on the long-lived qubit via transient excitation to these Rydberg states. Specifically, we apply a time-optimal gate protocol~\cite{Jandura2022} consisting of a parameterized sinusoidal phase modulation~\cite{Evered2023} and a constant Rydberg Rabi frequency with finite rise/fall time, limited by acousto-optical modulators (AOM) (Fig.~\ref{fig:overview}e). Up to experimental imperfections, this pulse introduces a CZ gate unitary and a single-atom phase~\cite{Levine2019} that is compensated with a calibrated virtual $Z(\phi)$ gate, which is a global phase shift on the clock laser AOM (Appendix~\ref{Appendix:SSB}). 

\subsection*{Gate fidelity metrics}
Before describing the benchmarking sequence, it is instructive to discuss and define different metrics of gate fidelity relevant in the context of benchmarking Rydberg-based CZ gates. Generally, the \textit{average fidelity}, or simply \textit{fidelity}, is a measure of how well an experimental quantum operation realizes an ideal unitary~\cite{Nielsen2002}. If not specified further, the convention is to average over input states drawn from the uniform (Haar) measure on the relevant Hilbert space. For the CZ gates discussed in this work, the average fidelity over two-qubit Haar random states is
\begin{equation}
    \begin{split}
        F_\mathrm{Haar}\equiv \mathbb{E}_{\ket{\psi}\sim \mathrm{ Haar}}\Tr \left[\mathrm{CZ} \ketbra{\psi}\mathrm{CZ}\ \mathcal{E}_\mathrm{CZ}(\ketbra{\psi}) \right]
    \end{split}
\end{equation}
where $\mathcal{E}_\mathrm{CZ}$ represents the experimentally implemented quantum channel and $\mathrm{CZ}$ the ideal CZ gate unitary ($\mathrm{CZ}^\dagger=\mathrm{CZ}$).

Motivated by recently realized benchmarking sequences for neutral atoms~\cite{Evered2023,Ma2023}, which utilize global symmetric operations, we also introduce a \textit{symmetric fidelity}, which is the average over all two-qubit symmetric (with respect to exchange of the two qubits) Haar random input states:
\begin{equation}
    \begin{split}
        F_\mathrm{Sym}\equiv \mathbb{E}_{\ket{\psi}\sim \mathrm{Sym\ Haar}}\Tr \left[\mathrm{CZ} \ketbra{\psi}\mathrm{CZ}\ \mathcal{E}_\mathrm{CZ}(\ketbra{\psi}) \right].
    \end{split}
\end{equation}
As shown below, these two metrics are generally not equal under typical noise channels present in neutral atom experiments. In particular, for gates with errors dominated by decay from transient Rydberg states, typically $F_\mathrm{Haar}\neq F_\mathrm{Sym}$.

\subsection*{Benchmarking sequences}
Our goal is to develop a benchmarking sequence that isolates the fidelity of native Rydberg-based CZ gates, allowing us to probe for error mechanisms~\cite{Evered2023, Ma2023, Finkelstein2024} in the CZ gate itself, such as laser noise and decay from excitation to Rydberg states. To this end, the sequence should be insensitive to other error sources, namely state preparation, measurement errors, and single-qubit gate errors.

To achieve such insensitivity, the common practice in the broader context of quantum computing is randomized benchmarking~\cite{Knill2007, Magesan2011} or its variants, such as interleaved randomized benchmarking (IRB)~\cite{Magesan2012}. The central idea of these sequences is to execute a quantum circuit of randomly chosen gates from a pre-defined set and fit the final state fidelity versus the number of applied gates to a model function. The experimental implementation of these sequences usually requires extensive local control capability to ensure sufficient randomization.

By contrast, a typical neutral atom array experiment utilizes global excitation beams, often paired with some degrees of local control~\cite{Graham2022, Lis2023A, Eckner2023, Shaw2024,Bluvstein2024}. In particular, two-qubit entangling gates can be applied in parallel over many pairs, using a shared Rydberg excitation beam, combined with control over atomic positions to modify gate connectivities~\cite{Bluvstein2022,Finkelstein2024}. Thus, it is of particular interest to devise a gate benchmarking sequence that requires only global control capabilities. Subspace randomized benchmarking (SRB)~\cite{Baldwin2020} provides a possible solution by restricting the circuits to global controls only. However, we are interested in characterizing only a single element of the Clifford group --- the CZ gate --- while SRB benchmarks the entire Clifford group.

In this context, a certain type of global randomized benchmarking sequence, inspired by IRB, has been proposed and implemented on neutral atom experiments~\cite{Evered2023,Ma2023}, where a variable number of CZ gates are interleaved with global single-qubit rotations to benchmark Rydberg-based CZ gates. The quoted gate fidelity is obtained by fitting the decay trend of the final circuit fidelity versus the number of applied CZ gates, $N_\mathrm{CZ}$. As we show below, this sequence provides a benchmark for $F_\mathrm{Sym}$, but has residual sensitivity to \textit{single-qubit gate errors} despite the fact that the number of applied single-qubit gates is kept constant (as discussed in Appendix~\ref{Appendix:SSB}, Fig.~\ref{SI_Harvard} and in our previous work~\cite{Finkelstein2024}). We interpret this sensitivity to stem from the fact that the distribution over two-qubit states varies as a function of $N_\mathrm{CZ}$. In particular, the time spent in a fully separable state varies with $N_\mathrm{CZ}$, leading to differences in sensitivity to single-qubit gate errors as a function of $N_\mathrm{CZ}$. 

This sensitivity to single-qubit gate errors is particularly relevant in our experiment. Due to laser noise on the clock laser, the single-qubit rotation fidelity of 0.9978(4) is comparable with the fidelity of the entangling operation~\cite{Finkelstein2024} (Appendix~\ref{Appendix:Setup}). Hence, we are interested in a benchmarking sequence that isolates the CZ fidelity itself with minimal sensitivity to single-qubit gate errors.

\subsection*{Symmetric stabilizer benchmarking (SSB)}
Inspired by IRB and the aforementioned sequences~\cite{Evered2023,Ma2023}, we design a new sequence that is straightforward to implement on experiments with global control (\textit{e.g.}, neutral atom experiments), yields a measure of the CZ fidelity in the subspace where qubit exchange symmetry is preserved, $F_\mathrm{Sym}$, and is largely insensitive to single-qubit gate errors.

Our circuit is particularly simple as it involves only global single-qubit $\pi/2$ rotations around randomly drawn axes $\pm X$ and $\pm Y$ $\{R_{\pm X}(\pi/2), R_{\pm Y}(\pi/2)\}$, interleaved with a different number of CZ gates (Fig.~\ref{SI_SSB_seq}a). Since the initial state also has qubit exchange symmetry, all states accessible during the circuit are twelve symmetric stabilizer states (SSS), each of which is described by two different stabilizers~\cite{Gottesman1997} $\pm \sigma_\mu \sigma_\nu, \pm \sigma_\nu \sigma_\mu$, where $\sigma_\mu, \sigma_\nu \in \{I, X, Y, Z\}$ (Table~\ref{tab:SSS}). 

\begingroup
\begin{table}[h!]
    \centering
    \renewcommand{\arraystretch}{1.15}
    \begin{tabular}{c|c}
    \hline
       Stabilizers  &  State \\
       \hline
    $IX, XI$ & $\frac{1}{2} \ket{00} + \frac{1}{2} \ket{01} + \frac{1}{2} \ket{10} + \frac{1}{2} \ket{11}$\\
    $-IX, -XI$ & $\frac{1}{2} \ket{00} - \frac{1}{2} \ket{01} - \frac{1}{2} \ket{10} + \frac{1}{2} \ket{11}$\\
    $IY, YI$  & $\frac{1}{2} \ket{00} + \frac{i}{2} \ket{01} + \frac{i}{2} \ket{10} - \frac{1}{2} \ket{11}$\\
    $-IY, -YI$ & $\frac{1}{2} \ket{00} - \frac{i}{2} \ket{01} - \frac{i}{2} \ket{10} - \frac{1}{2} \ket{11}$\\
    $IZ, ZI$ & $\ket{00}$\\
    $-IZ, -ZI$  & $\ket{11}$\\
    $XZ, ZX$ & $\frac{1}{2} \ket{00} + \frac{1}{2} \ket{01} + \frac{1}{2} \ket{10} - \frac{1}{2} \ket{11}$\\
    $-XZ, -ZX$ & $\frac{1}{2} \ket{00} - \frac{1}{2} \ket{01} - \frac{1}{2} \ket{10} - \frac{1}{2} \ket{11}$\\
    $YZ, ZY$ & $\frac{1}{2} \ket{00} + \frac{i}{2} \ket{01} + \frac{i}{2} \ket{10} + \frac{1}{2} \ket{11}$\\
    $-YZ, -ZY$ & $\frac{1}{2} \ket{00} - \frac{i}{2} \ket{01} - \frac{i}{2} \ket{10} + \frac{1}{2} \ket{11}$\\
    $XY, YX$ & $\frac{1}{\sqrt{2}} \ket{00} + \frac{i}{\sqrt{2}} \ket{11}$\\
    $-XY, -YX$ & $\frac{1}{\sqrt{2}} \ket{00} - \frac{i}{\sqrt{2}} \ket{11}$\\
    \hline
    \end{tabular}
    \caption{\textbf{The twelve symmetric stabilizer states.} The twelve symmetric stabilizer states with corresponding stabilizer operators (left column) are written out in the computational basis (right column).}
    \label{tab:SSS}
    \vspace{-0.0cm}
\end{table}
\endgroup

Importantly, these twelve SSS form a quantum state 2-design~\cite{Ambainis2007,Choi2023} on the symmetric subspace (which can be verified by explicitly computing the frame potential~\cite{Scott2008}); as a consequence, the gate fidelity averaged over the twelve SSS, $F_\mathrm{SSS}$, equals the symmetric fidelity, $F_\mathrm{Sym}$: 
\begin{equation}\label{equ:FSS}
    \begin{split}
        F_\mathrm{SSS} &\equiv \mathbb{E}_{\ket{\psi}\in \mathrm{SSS}}\Tr \left[\mathrm{CZ} \ketbra{\psi}\mathrm{CZ}\ \mathcal{E}_\mathrm{CZ}(\ketbra{\psi}) \right]\\
         &= F_\mathrm{Sym}\ .
    \end{split}
\end{equation}

Further, the sequence is designed to maintain a uniform distribution over these twelve SSS as the number of CZ gates, $N_\mathrm{CZ}$, is varied ($N_\mathrm{CZ}\geq2$, see below). An initialization unitary $\hat{U}_\mathrm{init}$ prepares the uniform distribution (Table~\ref{tab:Uinit}). This is followed by a varying number of CZ gates applied together with global single-qubit $\pi/2$ rotations, randomly drawn from $\{R_{\pm X}(\pi/2), R_{\pm Y}(\pi/2)\}$; importantly, the uniform distribution over SSS is unchanged under such operations. Hence, the sequence indeed maintains a uniform distribution over SSS independent of $N_\mathrm{CZ}$, in the absence of errors. As a consequence, a single-qubit gate error occurring at any time in the circuit, past $\hat{U}_\mathrm{init}$, acts on the same state distribution, independent of $N_\mathrm{CZ}$. Combined with the design that the number of single-qubit operations is held constant, we expect reduced sensitivity to single-qubit gate errors, including realistic error sources for our system (see below).

The sequence closes with a recovery unitary $\hat{U}_\mathrm{rec}$, which returns the final state to $\ket{11}$ in the absence of errors (Table~\ref{tab:Urec}). Both initialization $\hat{U}_\mathrm{init}$ and recovery $\hat{U}_\mathrm{rec}$ unitaries contain one CZ gate each and a fixed number of single-qubit rotations (Fig.~\ref{SI_SSB_seq}a). 

In the experiment, we vary $N_\mathrm{CZ}$ (including the two CZ gates in $\hat{U}_\mathrm{init}$ and $\hat{U}_\mathrm{rec}$) from two to ten and measure the probabilities that the final two-qubit state is $\ket{11}$ ($P_{\ket{11}}$). We fit the return probability $P_{\ket{11}}$ with the model $a_0\times F^{N_\mathrm{CZ}}$ and quote $F$ as the fidelity extracted from the symmetric stabilizer benchmarking sequence, $F_\mathrm{SSB}$ (Appendix~\ref{Appendix:Setup}). We note that this fit function gives a good estimate of $F_\mathrm{SSS}$ (which matches $F_\mathrm{Sym}$, see Eq.~\ref{equ:FSS}) under purely depolarizing noise up to linear order in the depolarizing noise strength. Under these specific noise assumptions, we have
\begin{equation}\label{Eq:SSB_SSS_Sym}
    F_\mathrm{SSB} \approx  F_\mathrm{Sym}\ .
\end{equation}

\begin{figure}[ht!]
    \centering
    \includegraphics[width=\columnwidth]{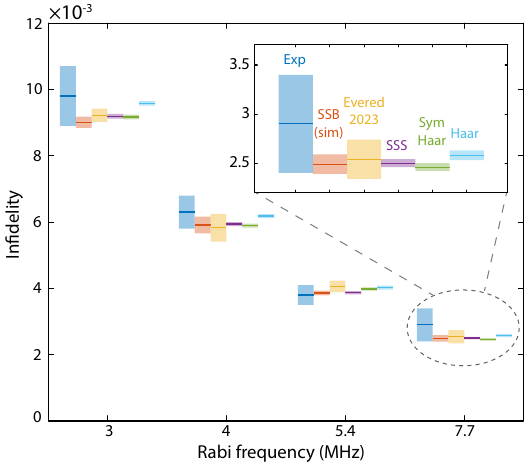}
    \caption{\textbf{Comparison of experimental results and various numerical metrics obtained from a full \textit{ab initio} error model simulation.} 
    Experimental results for the symmetric stabilizer benchmarking sequence as a function of Rabi frequency, $F_\mathrm{SSB}^\mathrm{exp}$ (Exp), together with full error model simulations of the same sequence, $F_\mathrm{SSB}^\mathrm{sim}$ (SSB (sim)), and the sequence of Ref~\cite{Evered2023} (Evered 2023). We also show simulations, with the same error model, for the gate fidelity averaged over symmetric stabilizer states, $F_\mathrm{SSS}$ (SSS), symmetric Haar random states, $F_\mathrm{Sym}$ (Sym Haar), and two-qubit Haar random states, $F_\mathrm{Haar}$ (Haar). Results from benchmarking sequences and these fidelities all agree within error bars with a notable exception being the fidelity over two-qubit Haar random states, as discussed in the text. Error bars on $F_\mathrm{SSB}^\mathrm{exp}$ and $F_\mathrm{SSB}^\mathrm{sim}$ are obtained from the maximum-likelihood fitting method (Appendix~\ref{Appendix:Setup}) with respective data points and their uncertainties. Error bars on simulation results (including individual data points for fitting $F_\mathrm{SSB}^\mathrm{sim}$) are statistical uncertainties from averaging over noisy trajectories (see Appendix~\ref{Appendix:Ryd} for error model simulation). Both represent one-sigma confidence intervals on the estimated mean. Numerical values are provided in Table~\ref{tab:numbers_comparison}.
    }
    \label{fig:comparison_benchmark}
    \vspace{-0.3cm}
\end{figure}

\begin{figure}[ht!]
    \centering
    \includegraphics[width=\columnwidth]{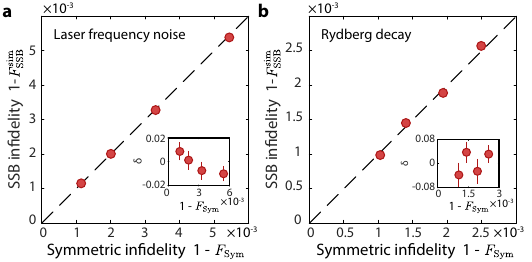}
    \caption{\textbf{Simulation of the circuit-inferred infidelity $1-F_\mathrm{SSB}^\mathrm{sim}$ under individual error sources.} 
    We simulate the SSB circuit, in the absence of single-qubit gate errors, at different Rydberg Rabi frequencies with each of the two major error sources, \textbf{(a)} laser frequency noise and \textbf{(b)} Rydberg decay, individually turned on. The dashed line is a guide to the eye for 1:1 ratio. The fractional difference $\delta = [(1 - F_\mathrm{SSB}^\mathrm{sim}) - (1 - F_\mathrm{Sym})] / (1-F_\mathrm{Sym})$ is plotted in the insets. The differences between $F_\mathrm{SSB}^\mathrm{sim}$ and $F_\mathrm{Sym}$ are below $1\times 10^{-4}$ for the Rabi frequencies used in the experiment.
    }
    \label{fig:comparison_noise_source}
    \vspace{-0.3cm}
\end{figure}

\subsection*{Numerical simulation with an \textit{ab initio} error model}
While the above relation holds for purely depolarizing noise and certain other noise channels (see Appendix~\ref{Appendix:SSB}3), on a generic experimental setup, the error channels are more complicated than pure depolarizing channels. In this case, our sequence, as well as many sequences in the literature~\cite{Knill2007,Magesan2011,Magesan2012,Baldwin2020,Evered2023,Ma2023}, does not have a theoretical guarantee. Hence, in this work we utilize a detailed error model to systematically compare the value inferred from our proposed circuit, $F_\mathrm{SSB}$, with the average gate fidelity metrics  $F_\mathrm{SSS}$, $F_\mathrm{Sym}$, and $F_\mathrm{Haar}$.

To this end, we utilize an \textit{ab initio} Rydberg error model simulation~\cite{Shaw2024B} used in our previous works. The error model includes four dominant error sources: laser frequency noise, laser intensity noise, spontaneous and blackbody-radiation-induced decay, and atomic motion (Appendix~\ref{Appendix:Ryd}). All sources are independently calibrated, \textit{e.g.}, we measured the power spectral density of frequency and intensity noise~\cite{Shaw2024B} as well as the decay rates from Rydberg states~\cite{Scholl2023A} separately. Previously, the model was also tested against the fidelity for generating Bell states in a meta-stable-state to Rydberg-state qubit~\cite{Scholl2023A} and for generating large-scale entangled states via many-body quench dynamics~\cite{Choi2023,Shaw2024C}. We provide a further test below, specifically isolating the power spectral density of fast frequency noise, utilizing a spin-lock sequence~\cite{Finkelstein2024} (Section~\ref{Sec:application} and Fig.~\ref{fig:spin_lock}).

In the context of CZ fidelities, we first test if the error model reproduces the experimental results of the SSB sequence for different gate speeds and consequently with varying contributions of different error sources (Fig.~\ref{fig:overview}f). To this end, we repeat the SSB sequence for gates with different Rabi frequencies and extract the experimental values of $F_\mathrm{SSB}^\mathrm{exp}$. The Rabi frequency determines the evolution speed under the applied Rydberg pulse and thus the gate duration, which naturally varies the strength of each error source. For example, Rydberg decay becomes more dominant for small Rabi frequencies as the gate duration gets longer. We then simulate the sequence with the \textit{ab initio} error model and obtain $F_\mathrm{SSB}^\mathrm{sim}$. We find good agreement between $F_\mathrm{SSB}^\mathrm{sim}$ and $F_\mathrm{SSB}^\mathrm{exp}$ within experimental error bars across a range of Rabi frequencies [Exp (blue) and SSB (sim) (red) in Fig.~\ref{fig:comparison_benchmark}].

We further utilize this tested error model to check if the benchmarking sequence indeed yields the average gate fidelity over symmetric input states, $F_\mathrm{Sym}$ (Eq.~\ref{Eq:SSB_SSS_Sym}). First, we find that the simulation of the benchmarking sequence, $F_\mathrm{SSB}^\mathrm{sim}$, is consistent with the numerical calculation of gate fidelity $F_\mathrm{Sym}$ for different Rydberg Rabi frequencies [SSB (sim) (red) and Sym Haar (green) in Fig.~\ref{fig:comparison_benchmark}]. 
Besides, we identify that in the range of Rabi frequencies we consider, the two major error sources are laser frequency noise and Rydberg decay (Fig.~\ref{fig:overview}c,f). 

Further, we simulate the fidelity prediction of SSB, $F^\mathrm{sim}_\mathrm{SSB}$, and find that it differs from $F_\mathrm{Sym}$ by no more than $1\times 10^{-4}$, well below our experimental uncertainties, for these dominant error sources \textit{individually} at all Rydberg Rabi frequencies used in experiment (Fig.~\ref{fig:comparison_noise_source}).
Hence, our sequence yields the CZ gate fidelity averaged over the symmetric Haar subspace under realistic noise sources: 
\begin{equation}
    F_\mathrm{SSB} \approx F_\mathrm{Sym}\ \mathrm{(realistic\ noise\ sources)}.
\end{equation}

We note that $F_\mathrm{SSS}$ [SSS (purple) in Fig.~\ref{fig:comparison_benchmark}] also agrees with $F_\mathrm{Sym}$ within numerical error bars, which serves as a numerical check of the equality stated in Eq.~\ref{equ:FSS} with our error model.

In addition, we simulate (with our error model) the sequence presented in Ext. Data Fig. 6 in Ref.~\cite{Evered2023} [Evered 2023 (yellow) in Fig.~\ref{fig:comparison_benchmark}] and find good agreement with all aforementioned metrics in the absence of single-qubit gate errors. Hence, the benchmarking results from our work can be compared with benchmarking values presented in other recent works~\cite{Evered2023,Ma2023}. Further, our simulation also shows that the underlying fidelity metric targeted in these works is $F_\mathrm{Sym}$. 

\subsection*{Reduced sensitivity to single-qubit gate errors}

\begin{figure}[ht!]
    \centering
    \includegraphics[width=\columnwidth]{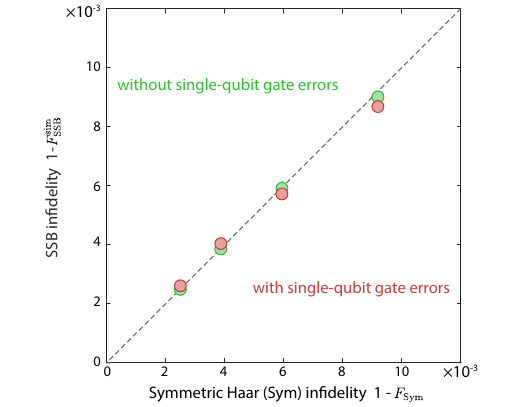}
    \caption{\textbf{Circuit-inferred fidelity $F_\mathrm{SSB}^\mathrm{sim}$ with and without single-qubit gate errors.} We simulate, in our \textit{ab initio} error model, the SSB sequence with and without single-qubit gate errors at the four Rydberg Rabi frequencies used in the experiment. In the absence of single-qubit gate errors, our numerical simulation shows that the circuit infers a good proxy of the fidelity averaged over symmetric Haar random states $F_\mathrm{Sym}$ (green markers). Including single-qubit gate errors in the clock error model (Appendix~\ref{Appendix:Clock}), we find the sequence still yields infidelity consistent within the experimental uncertainty (red markers). Dashed line is a guide to the eye for 1:1 ratio.}
    \vspace{-0.2cm}
    \label{SI_corr_clock}
\end{figure}

While single-qubit gate errors are not included in the simulations in Fig.~\ref{fig:comparison_benchmark}, we want to emphasize that even in the presence of single-qubit gate errors, our sequence remains a good proxy of $F_\mathrm{Sym}$, which we confirm with a detailed numerical simulation including realistic single-qubit noise channels (Fig.~\ref{SI_corr_clock}). We find that the effect of single-qubit gate errors is below the experimental uncertainty (Fig.~\ref{SI_corr_clock}, red markers). This is in contrast to the sequence from Evered 2023 (Fig.~\ref{SI_Harvard}). We suggest a modification of this sequence to be less sensitive to single-qubit gate errors in Appendix~\ref{Appendix:SSB}, where we also show several analytical models. 

Finally, we note that in principle one can construct an exotic error channel (\textit{e.g.}, strong coherent errors in both single-qubit and two-qubit operations) and find this benchmarking method to fail. However, this is likely unrealistic and not the case for our experimental system: we measure separately the shot-to-shot laser intensity and frequency noise and simulate the worst-case scenario -- all CZ gates share the same shot-to-shot noise in the circuit for a single realization, which leads to strongly coherent errors. The simulation results are shown in Table~\ref{tab:infidelity DC noise}. We find that even at the smallest Rydberg Rabi frequency, the circuit prediction is only off by $1\times 10^{-4}$.
We also note, in particular, that not only our Rydberg error model but also our single-qubit gate error model has been extensively tested~\cite{Finkelstein2024}. In addition, to the best of our knowledge, there is no benchmarking circuit that can circumvent this issue, including randomized benchmarking~\cite{Figueroa-Romero2021}.

\subsection*{Comparison between $F_\mathrm{Sym}$ and $F_\mathrm{Haar}$}
We also compare the fidelity averaged over symmetric Haar random states, $F_\mathrm{Sym}$, and that averaged over Haar random states, $F_\mathrm{Haar}$ [Sym Haar (green) and Haar (light blue) in Fig.~\ref{fig:comparison_benchmark}]. As the Rydberg pulse gets longer (decreasing Rabi frequency), the growing inconsistency between the two indicates that the benchmarking sequence presented in this work, as well as those in other works that only involve global operations, are not benchmarking the fidelity of the CZ gate averaged over two-qubit Haar random states, $F_\mathrm{Haar}$.

A possible explanation comes from the fact that the decay probabilities over these two sets of input states are different. As a straightforward illustration, assuming the decay probability from the transient Rydberg state excitation during the CZ gate for the initial state $\ket{01}$ is $\varepsilon_1$ and that for the initial state $\ket{11}$ is $\varepsilon_2$, then the average decay probability for two-qubit Haar random states is $\varepsilon^d_\text{Haar}=(2\varepsilon_1 + \varepsilon_2)/4$ while that for two-qubit symmetric Haar random states is $\varepsilon^d_\text{Sym}=(\varepsilon_1 + \varepsilon_2)/3$. Hence, the Rydberg-decay-induced infidelity is different for these two sets of input state. Note though that $\varepsilon^d_\text{Haar}$ can be bounded by $\varepsilon^d_\text{Sym}$. In particular,
\begin{equation}
    \dfrac{\varepsilon^d_\text{Haar}}{\varepsilon^d_\text{Sym}} = \dfrac{3(2\varepsilon_1 + \varepsilon_2)}{4(\varepsilon_1 + \varepsilon_2)} \in \left(\dfrac{3}{4}, \dfrac{3}{2}\right).
\end{equation}
Hence, in the worst case scenario, the infidelity due to decay when averaging over full two-qubit Hilbert space is $50\%$ higher than that averaging over symmetric states. Note, in our specific case, the ratio is significantly smaller than this worst case scenario: $\varepsilon^d_\text{Haar}/\varepsilon^d_\text{Sym}=2.9/2.6 \approx 1.12 $. We refer readers to Appendix~\ref{Appendix:CompareCZ} for more details and a discussion of other error sources.

With these comparisons, we confirm that under a wide range of realistic experimental imperfections, $F^{\text{exp}}_{\text{SSB}}\approx F_{\text{Sym}}\neq F_{\text{Haar}}$. Note though that for the highest Rabi frequency, $F_{\text{Haar}}$ is within error bars of the simulated, benchmarked fidelity $F^{\text{sim}}_{\text{SSB}}$. This stems from the fact that for large Rabi frequencies, laser noise errors dominate, for which the differences between $F_{\text{Haar}}$ and $F_{\text{Sym}}$ are small (Appendix~\ref{Appendix:CompareCZ}).
\bigskip

In summary, we have developed a new benchmarking sequence --- symmetric stabilizer benchmarking --- that isolates the symmetric CZ gate infidelity and has reduced sensitivity to single-qubit gate errors. With a tested \textit{ab initio} error model, we confirm that this sequence provides a measure of $F_\mathrm{Sym}$, the CZ gate fidelity averaged over the two-qubit symmetric Haar states. On our experiment, we correct for false contribution from leakage (Appendix~\ref{Appendix:BrightState}) and find a fidelity of \highFidBBCorrected\ at our current highest attainable Rydberg Rabi frequency of \highRabi\ (Fig.~\ref{fig:overview}b). This number, together with our recent work~\cite{Finkelstein2024}, forms a new state-of-the-art among neutral atom array experiments.

\section{Fidelity response theory (FRT) applied to modeling errors}\label{Sec:linear_response}
The full \textit{ab initio} error model simulation quantitatively predicts the CZ gate fidelity given the laser noise spectra and the pulse profile. Yet, such full numerical models are computational-resource-demanding. In addition, we would like to gain analytical insight and a more intuitive understanding of how each error source affects the gate fidelity and how different gate protocols (\textit{e.g.} Refs~\cite{Levine2019, Fromonteil2023}) would perform on our experiment. To this end, we develop FRT that predicts the infidelity of any quantum process under any given noise PSD that qualifies as a small perturbation to the system~\cite{Poggi2024}.

The infidelity of a gate operation, due to laser intensity and frequency noise, can be understood from the perspective of a perturbation on the driving Hamiltonian. Previously, this approach has been applied to analyzing the performance of several platform-specific operations ~\cite{Jiang2023,deLeseleuc2018,Nakav2023,Ball2015,Green2013,Day2022}. We generalize this framework to any unitary quantum processes with the following Hamiltonian,
\begin{equation}
    \hat{H}(t) = \hat{H}_0(t) + \sum_j h_j(t) \hat{O}_j(t),
\end{equation}
where $\hat{H}_0(t)$ is the target, programmed time-dependent drive, and each $h_j(t) \hat{O}_j(t)$ is an independent noise term on the Hamiltonian, with $h_j(t)$ described by a power spectral density $S_j(f)$, and $\hat{O}_j(t)$ a noise operator that only depends on the noise type and is independent of the noise spectrum.

Assuming that all noise sources are described by mutually uncorrelated PSDs and the error rate is low, the infidelity can be expressed as a linear combination of PSDs weighted with fidelity response functions. Specifically, the infidelity of a certain quantum state evolving under the Hamiltonian is a linear function in the power spectral density to leading order,
\begin{equation}\label{eq:sum_error_sources}
    1 - F \approx \sum_j \int df S_j(f) I_j(f)
\end{equation}
where $I_j(f)$ is the linear infidelity response function of noise $j$, in the sense that the induced infidelity is linear in the noise power spectral density $S_j(f)$. The total infidelity is the sum of infidelity contributions from each noise channel. 

More concretely, the fidelity response function $I(f)$ is the ratio of the induced infidelity to the strength of a delta function PSD at frequency $f$. This can be seen by assuming only one noise channel with $S(f)=\epsilon\delta(f_0-f)$.  The infidelity is then
$1-F=\epsilon I(f_0) + O(\epsilon^2)$. Hence, 
$I(f_0)=\lim_{\epsilon \rightarrow 0} (1-F)/\epsilon$. 
This approach can be used to numerically find the response functions by simulating the quantum process with a delta function noise term for various frequencies and for different noise channels. In addition, one can experimentally determine the response functions by injecting a single-frequency modulation and measuring the decrease in fidelity. As an example, we demonstrate experimentally measuring the intensity noise response of the CZ gate in Appendix~\ref{Appendix:exp_response_function}.

Alternatively, we provide a closed-form expression for evaluating the response function. For a noise operator $\hat{O}(t)$, the response function is analytically found to be (see Appendix~\ref{Appendix:LinearResponse} for derivation)
\begin{equation}\label{eq:response_func}
    I(f) = \int_0^T dt \int_0^T d\tau \cos (2\pi f(t-\tau)) \langle \hat{O}_H(t) \hat{O}_H(\tau) \rangle_c
\end{equation}
where $\hat{O}_H(t) = \hat{U}^\dagger (t) \hat{O}(t) \hat{U}(t)$ is the noise operator in the Heisenberg picture, $\langle \hat{O}_H(t) \hat{O}_H(\tau) \rangle_c = \langle \hat{O}_H(t) \hat{O}_H(\tau) \rangle - \langle \hat{O}_H(t) \rangle \langle \hat{O}_H(\tau) \rangle$ is the connected correlator (evaluated with the initial state $\ket{\psi(t=0)}$), $\hat{U}(t)$ is the unitary evolution under $\hat{H}_0(t)$, and $T$ is the total evolution time. Throughout this work, we use this closed-form expression for numerical response function calculations.

\begin{figure*}[ht!]
    \centering
    \includegraphics[width=\textwidth]{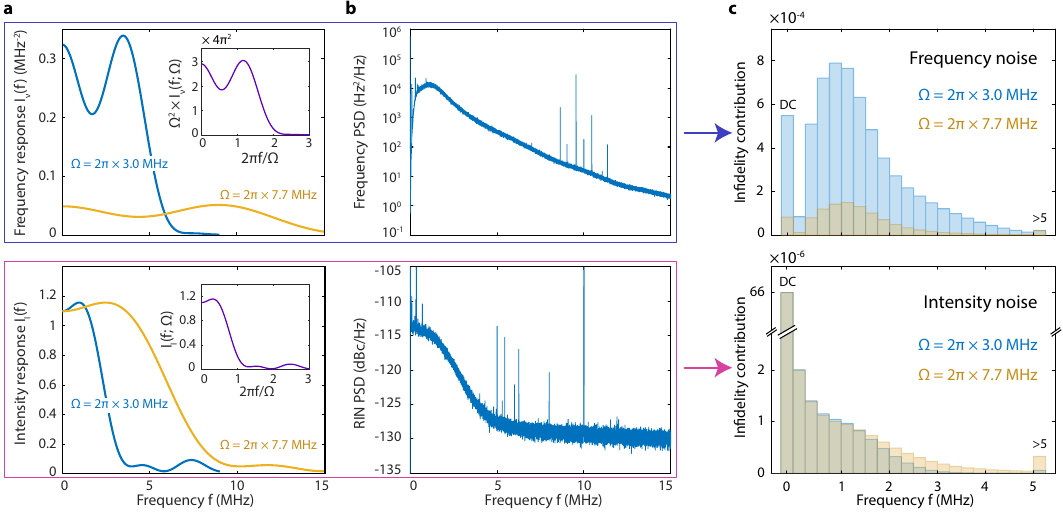}
    \caption{\textbf{Fidelity response of CZ gate to laser noise.} Upper panel of each subfigure corresponds to the Rydberg laser frequency noise, and the lower panel corresponds to the intensity noise.
    \textbf{(a)} Fidelity response functions for frequency noise and intensity noise with the time-optimal gate and the fidelity metric $F_\mathrm{Haar}$. For an \textit{ideal} time-optimal gate assuming zero rise/fall time, infinite blockade interaction strength, and absence of self-light shift, the response functions at different Rabi frequencies are directly related by a scaling law (see text), from which a universal response function (inset) is defined. To recover response functions at different Rabi frequencies, the universal response functions are rescaled along the frequency (horizontal) axis by $\Omega$, and the strength of frequency response (vertical) is rescaled by $\Omega^{-2}$ while that of intensity response does not scale. Comparison with the response functions averaged over symmetric Haar random states, which we use to predict experimental infidelity throughout this work, is depicted in Fig.~\ref{SI_response_compare}.
    \textbf{(b)} Laser frequency noise and relative intensity noise (RIN) power spectral densities (PSD), measured by an in-loop PDH error signal, and a photodiode, respectively. Shot-to-shot (DC) noise is not shown explicitly here but is taken into account for a complete description of the laser noise.
    \textbf{(c)} Taking the product of the response function at a given Rabi frequency and the corresponding PSD, we compute the infidelity contribution density. Here, we integrate the resulting curve over segments of 250~kHz to illustrate the contributions from each range of the laser noise. For illustration purposes, we separate out contributions from the low frequencies (DC) and contributions greater than 5~MHz. We show the resulting histogram for Rabi frequency at 3~MHz and 7.7~MHz.
    }
    \vspace{-0.3cm}
    \label{fig:linear_response}
\end{figure*}

As an instructive example, we numerically calculate response functions to laser intensity noise and laser frequency noise in a simplified model for an \textit{ideal} time-optimal CZ gate with zero rise/fall time, infinite blockade interaction strength, and absence of self-light shift. The simplified model is detailed in the next subsection. We use the fidelity metric $F_\mathrm{Haar}$ to calculate the response functions for frequency and intensity noise shown in Fig.~\ref{fig:linear_response}a  (approximate functional forms are provided in Appendix~\ref{Appendix:ApproxForms}). The response functions for $F_\mathrm{Haar}$ are very similar to the ones for $F_\mathrm{Sym}$, the most relevant metric in Section~\ref{Sec:benchmarking} (Fig.~\ref{SI_response_compare}).

To study the infidelity due to frequency noise (intensity noise) at a given Rydberg Rabi frequency $\Omega/2\pi$, we multiply the laser frequency noise PSD $S_\nu(f)$ (laser relative intensity noise PSD $S_I(f)$) (Fig.~\ref{fig:linear_response}) by the corresponding frequency (intensity) response function $I_\nu (f; \Omega)$ ($I_I(f; \Omega)$). This yields the infidelity contribution density as a function of noise frequency (Fig.~\ref{fig:linear_response}c) at a given gate speed. For instance, we observe that the servo-induced noise~\cite{Li_Covey2022,Levine2019,Nakav2023} around 1~MHz contributes the most in the infidelity contribution due to frequency noise. On the other hand, the infidelity contribution due to intensity noise mainly comes from shot-to-shot, low-frequency noise (DC).

Finally, we integrate the contribution density to obtain the total infidelity $\varepsilon_\nu$  ($\varepsilon_I$) due to laser frequency (intensity) noise: 
\begin{align}
    \label{eq:epsilon_nu}
    \varepsilon_\nu &=  \int df S_\nu(f) I_\nu (f; \Omega)\ ,\\
    \varepsilon_I &=  \int df S_I(f) I_I (f; \Omega)\ .
    \label{eq:epsilon_I}
\end{align}

\subsection*{Universal scaling laws of response functions}
Interestingly, the response functions follow universal scaling laws with Rabi frequency. For example, for frequency noise, the width of the response function $I_\nu (f; \Omega)$ scales linearly with $\Omega$, while the sensitivity to slow noise decreases as $\Omega$ increases (Fig.~\ref{fig:linear_response}a, upper). For the intensity response $I_I(f; \Omega)$, the width also scales linearly with $\Omega$, but the sensitivity to slow noise remains unchanged (Fig.~\ref{fig:linear_response}a, lower). 

We now show analytically that the response functions for different Rabi frequencies indeed collapse after appropriate rescaling (Fig.~\ref{fig:linear_response}a, inset). This collapse holds under simplified model assumptions (zero rise/fall time, infinite blockade interaction strength, and absence of self-light shift). The simplified model consists of a Hamiltonian describing a two-qubit system under an ideal time-optimal gate~\cite{Jandura2022,Evered2023} (with $\hbar=1$)
\begin{equation}\label{eq:gateHamiltonian}
\begin{split}
     \hat{H}_0(t) = &\dfrac{\Omega}{2}\sum_{i=1}^2\left[e^{-i\varphi(t)} \ket{1_i}\bra{r_i} + e^{i\varphi(t)} \ket{r_i}\bra{1_i}\right] \\ &+ B\ketbra{rr},   
\end{split}
\end{equation}
where $\varphi(t)$ is the phase modulation, and $B$ is the blockade interaction. We take $B\rightarrow\infty$ in this simplified model.  

To calculate the response function (Eq.~\ref{eq:response_func}), we identify the frequency noise operator, $\hat{O}_\nu$, and the intensity noise operator, $\hat{O}_I$,
\begin{align}\label{equ:noise operators}
    \hat{O}_\nu(t) &= -2\pi \sum_{i=1}^2 \ket{r_i}\bra{r_i} \equiv \Tilde{O}_\nu(s), \\
    \begin{split}
    \hat{O}_I(t) &= \dfrac{\Omega}{4}\sum_{i=1}^2\left[e^{-i\varphi(t)} \ket{1_i}\bra{r_i} + e^{i\varphi(t)} \ket{r_i}\bra{1_i}\right]\\
    &\equiv \Omega \Tilde{O}_I(s).
    \end{split} 
\end{align}

These noise operators themselves can be rescaled into a universal form. To this end, we parameterize the phase modulation $\varphi(t)$ with normalized time $s = t/T$, where $T$ is the length of the pulse, which is inversely proportional to Rabi frequency $\Omega/2\pi$. We also separate out the Rabi-frequency-dependent part from intensity noise operator $\hat{O}_I$. This yields dimensionless operators $\Tilde{O}_\nu(s)$ and $\Tilde{O}_I(s)$ that are independent of the applied Rabi frequency. 

As a consequence, the response functions themselves follow a universal form. Plugging the dimensionless noise operators into Eq.~\ref{eq:response_func}, the response functions are expressed as a Rabi-frequency-dependent part multiplied by a dimensionless part, the latter of which is a universal function of the particular gate protocol chosen. Reorganizing the resulting expressions and isolating the universal part, we retrieve the scaling laws for frequency noise response $I_\nu$ and intensity noise response $I_I$:
\begin{align}\label{Eq:nu_scaling}
    I_\nu (f;\Omega) &= \Omega^{-2} g_\nu \left(\dfrac{2\pi f}{\Omega}\right), \\
    I_I (f;\Omega) &= g_I \left(\dfrac{2\pi f}{\Omega}\right), \label{Eq:I_Scaling}
\end{align}
where $g_\nu$ and $g_I$ are universal response functions for the time-optimal gate (Fig.~\ref{fig:linear_response}a inset). This shows that after appropriate scaling, the response functions of both noise sources collapse to these universal functions for all Rabi frequencies, as seen in the inset of Fig.~\ref{fig:linear_response}a.

\subsection*{Approximate power-law behavior of infidelity contributions}
These rescaling behaviors of response functions lead to approximate power-law dependence of the resulting infidelities on Rabi frequency. As an observation, both frequency and intensity response functions are roughly flat over the frequency interval up to the applied Rabi frequency $\Omega/2\pi$ (Fig.~\ref{fig:linear_response}a). We also observe that in our Rydberg laser, both frequency and intensity noise PSDs are largely contained within the lowest Rydberg Rabi frequency considered, \textit{i.e.} 3~MHz in this work (Fig.~\ref{fig:linear_response}b). These two observations suggest that the primary components in the noise PSDs contributing to infidelity are contained in the interval where the response functions are approximately constant.

\begin{figure}[ht!]
    \centering
    \includegraphics[width=\columnwidth]{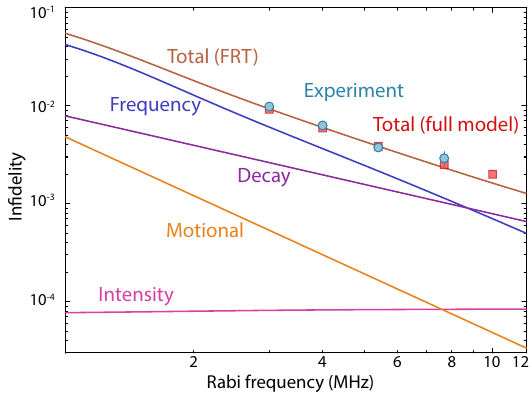}
    \caption{\textbf{The approximate power-law infidelity due to each error source as a function of Rydberg Rabi frequency.} This figure is a log-log version of Fig.~\ref{fig:overview}f which shows the approximate power-law behavior for all error sources. For each Rabi frequency $\Omega/2\pi$, following the procedure explained in Fig.~\ref{fig:linear_response} and taking the symmetric Haar fidelity response for \textit{ideal} time-optimal gates in Fig.~\ref{SI_response_compare}, we calculate the infidelity contribution from laser frequency noise and laser intensity noise separately. Both frequency and intensity noise contributions as a function of $\Omega$ exhibit approximate power-law dependence (see text). For the Rydberg decay, the error probability is proportional to the average time spent in the Rydberg state, yielding a $\Omega^{-1}$ dependence. The atomic motion in a tweezer trap leads to a velocity distribution, which induces a shot-to-shot frequency noise. Hence, its contribution scales as $\Omega^{-2}$. We take the total predicted infidelity and compare it against the experimental data and full \textit{ab initio} error model simulation with \textit{realistic} gates (red squares). The discrepancy between the full error model and the FRT prediction at high Rabi frequencies mainly stems from the finite rise and fall time of the AOM, which is not considered for \textit{ideal} gates (see text).
    }
    \vspace{-0.4cm}
    \label{fig:error_scaling}
\end{figure}

Further, the value of the flat part scales as $\Omega^{-2}$ for frequency noise response functions (Fig.~\ref{fig:linear_response}a, upper and Eq.~\ref{Eq:nu_scaling}). Hence, with the observation that the frequency noise PSD is largely contained in the same interval as the flat part of the response function, the total infidelity contribution from frequency noise is expected to scale roughly as $\Omega^{-2}$.

For intensity noise response functions, the value of the approximately flat part is constant with $\Omega$ (Fig.~\ref{fig:linear_response}a, lower and Eq.~\ref{Eq:I_Scaling}). Hence, this approximate model predicts a constant contribution to infidelity with Rabi frequency $\propto \Omega^0$.

We check these qualitative predictions with full FRT calculations (Eq.~\ref{eq:epsilon_nu} and Eq.~\ref{eq:epsilon_I}). First, we compute the response functions averaged over symmetric Haar states (Fig.~\ref{SI_response_compare}, approximate functional forms are provided in Appendix~\ref{Appendix:ApproxForms}). With our specific laser PSDs, we then calculate the infidelity contributions to $F_\mathrm{Sym}$ from laser frequency and intensity noise (blue and pink curves in Fig.~\ref{fig:error_scaling}). We see approximate power-law behaviors with fitted trend of $\Omega^{-1.791(2)}$ for frequency-noise-induced infidelity in the range of $\Omega > 2\pi\times 3~\text{MHz}$ while intensity-noise-induced infidelity converges increasingly to a constant asymptotically over the same range.

Other error sources also have power-law contributions to the total infidelity. The Rydberg decay probability is proportional to the gate duration, which is proportional to $\Omega^{-1}$ (purple curve in Fig.~\ref{fig:error_scaling}). The effect of atomic motion is dominated by the shot-to-shot Doppler-induced detuning. Hence, it can be viewed as slow frequency noise, whose contribution scales as $\Omega^{-2}$ (orange curve in Fig.~\ref{fig:error_scaling}). Summing up these four major contributions to the infidelity $1-F_\mathrm{Sym}$, we obtain a decreasing total infidelity as a function of Rabi frequency (brown curve in Fig.~\ref{fig:error_scaling}).

We also perform full \textit{ab initio} error model simulation of a \textit{realistic} time-optimal gate (with non-zero rise/fall time, finite blockade interaction strength, and self-light shift) over five different Rabi frequencies (Fig.~\ref{fig:error_scaling} red markers). Over the Rabi frequency range, the simulation agrees with the prediction from rescaled response functions, with slight deviation at higher Rabi frequencies mainly due to the effect of finite rise time. FRT can be easily applied to include finite rise time (Appendix~\ref{Appendix:realistic_gate}), but the universal scaling laws presented above would need to include finite correction terms for this effect.

\section{Applications of FRT}\label{Sec:application}

The proposed FRT is not only useful for predicting fidelities associated with CZ gates mediated by a one-photon Rydberg transition, but also for a wider range of applications as we illustrate in the following.

\medskip
\textit{Generalization to two-photon Rydberg transition.} --- FRT can be easily generalized to predict the fidelity for a CZ gate mediated by a two-photon Rydberg transition, as typically used in alkali atomic qubits~\cite{Levine2019,Graham2019}, by simulating the dynamics of a 4-level system. Alternatively, in the limit where the detuning to the intermediate state is much greater than the Rabi frequency of each single arm, the system can be simplified as a 3-level system with an effective Rabi frequency that depends on both lasers. The laser noise on each arm can be directly translated to effective noise on the effective one-photon transition (Appendix~\ref{Appendix:two_photon}).

\begin{figure}[ht!]
    \centering
    \includegraphics[width=\columnwidth]{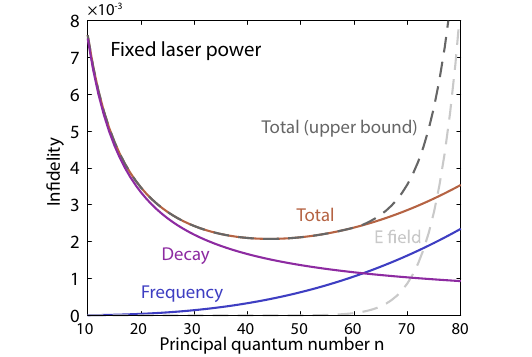}
    \caption{\textbf{CZ gate infidelity dependence on the Rydberg state principal quantum number with fixed laser power and laser noise spectrum.} Using FRT and Rydberg atom properties' dependence on principal quantum number $n$~\cite{Saffman2010}, we study the predicted time-optimal CZ gate infidelity as a function of $n$ at fixed laser power, corresponding to Rabi frequency of 7.7 MHz at $n=61$. We assume all detuning noise stems from laser frequency noise but not electric field noise. We apply FRT to predict the infidelity due to laser noise and rescale the Rydberg lifetime to predict the infidelity due to Rydberg decay. Individual contributions from the Rydberg decay and the laser frequency noise exhibit approximate power-law dependence in $n$. The total infidelity further includes laser intensity noise and finite atomic temperature. For this plot, we assume infinite blockade energy. However, for lower $n$, the true blockade energy is reduced. To maintain a fixed blockade-to-Rabi-frequency ratio between $n=61$ (the principle quantum number used for our experiment) and the predicted optimal quantum number, $n=44$, the inter-atomic distance would need to be modified from 3.3~$\mu\text{m}$ (for $n=61$) to 1.7~$\mu\text{m}$ (for $n=44$). Given the measured shot-to-shot detuning fluctuation on the atoms, we also infer an upper bound of the electric field noise, from which we calculate the upper bounds of infidelity due to electric field noise (light gray, dashed) and total infidelity (gray, dashed). The electric field noise does not alter the plot for $n\le 61$ but causes the infidelity to increase much faster for $n>61$.
    }
    \vspace{-0.3cm}
    \label{fig:n_dependence}
\end{figure}

\medskip
\textit{Comparing different Rydberg quantum states.} --- We study the CZ gate infidelity dependence on the principal quantum number $n$ of the Rydberg state~\cite{Saffman2010}. For larger $n$, the Rydberg state has longer lifetime, at the cost of larger sensitivity to electric field noise and lower Rabi frequency, which makes it more prone to laser frequency noise. Assuming a fixed laser power and the absence of electric field noise, we find minimal infidelity at $n=44$ (Fig.~\ref{fig:n_dependence}), which is notably smaller than the one in use currently in our experiment ($n=61$). In this plot, we assume an \textit{ideal} time-optimal gate (zero rise/fall time) and infinite Rydberg blockade energy ($B$ in the simplified model of Eq.~\ref{eq:gateHamiltonian}); however, for lower $n$, this approximation might not hold due to the reduced blockade energy and the increased Rabi frequency. Our experimental blockade-to-Rabi-frequency ratio at current $n$ is ${\approx}16$~\cite{Finkelstein2024} for which infinite $B$ is a good approximation. To maintain the same ratio, the interatomic distance $r\propto n^{25/12}$~\cite{Madjarovthesis} would need to be decreased to $1.7~\mu\text{m}$ for $n=44$, which is still accessible~\cite{Bluvstein2024}. As to finite rise/fall time, the Rabi frequency at $n=44$ would be 13~MHz, likely requiring an electro-optic modulator instead of an AOM to switch fast enough. In the case of non-zero electric field noise, a smaller $n$ than the current choice will also be preferable~\cite{Saffman2010}.

\begin{figure}[ht!]
    \centering
    \includegraphics[width=\columnwidth]{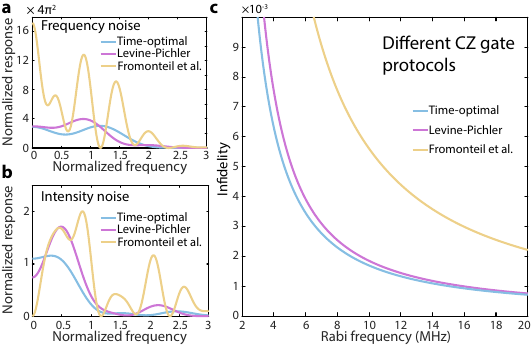}
    \caption{\textbf{Performance of different CZ gate protocols from FRT prediction.} We explore other gate protocols, including the Levine-Pichler gate~\cite{Levine2019} and the DC-intensity-noise-robust gate (Fromonteil et al.)~\cite{Fromonteil2023}, under our specific laser noise. Here we assume no rise/fall time effect and no self-light shift.
    \textbf{(a)} The laser frequency noise response comparison among the three considered gate protocols.
    \textbf{(b)} The laser intensity noise response comparison among the three considered gate protocols. Notably, the gate proposed by Fromonteil et al. is designed to be insensitive to static laser intensity noise, which shows up as a vanishing response at zero frequency.
    \textbf{(c)} Total CZ gate infidelity, including errors from laser noise, Rydberg decay, and atomic motion, as a function of maximal applied Rabi frequency during the pulse for the three considered protocols. Under our specific error sources, the time-optimal gate protocol appears to perform well, compared to other gate protocols studied.
    }
    \vspace{-0.2cm}
    \label{fig:different_gates}
\end{figure}
\medskip
\textit{Comparing gate protocols.} --- Our analytic approach to predicting gate fidelity allows one to choose and optimize the gate protocol under specific error sources. Using FRT, we compute the fidelities for different CZ gate protocols (Fig.~\ref{fig:different_gates}) and find that the time-optimal gate~\cite{Jandura2022,Evered2023} would yield the highest fidelity on our experiment, among the gate protocols studied, for all Rabi frequencies. We also note that the gate which minimizes time spent in Rydberg states~\cite{Pagano2022} is almost indistinguishable from the time-optimal gate, as the time spent in Rydberg states is only ${\approx} 0.3\%$ shorter than the time-optimal gate. Hence, we expect a similar performance for this gate.

We have also tried to optimize the pulse with the chopped random basis (CRAB) algorithm~\cite{Muller2022} and find that the time-optimal gate is near-optimal under our error model. Our analytic framework enables optimal control methods~\cite{Muller2022, Caneva2011} to be applied to any quantum process for more complicated error models beyond static noise and white noise, which were typically considered~\cite{Aroch2023}.

\begin{figure}[ht!]
    \centering
    \includegraphics[width=\columnwidth]{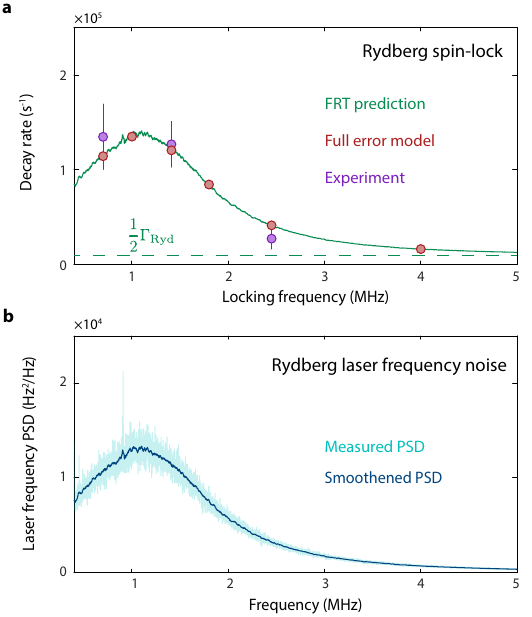}
    \caption{\textbf{Rydberg spin-lock.}
    \textbf{(a)} Comparison between the analytical prediction (dark green, obtained from FRT), the full error model simulation (red data points), and the experimental data points (purple data points with error bars) of the decay rate in a Rydberg spin-lock experiment (see text and Appendix~\ref{Appendix:spin_lock} for more details on this sequence). FRT predicts the spin-lock decay to be proportional to the smoothened frequency noise PSD (shown in b) up to an offset from the finite Rydberg lifetime (green dashed line). The agreement between numerical data points and the analytical curve justifies the use of FRT in our experimental noise regime for this experiment. The agreement between experimental results and numerical results further serves as a test of our laser frequency noise PSD.
    \textbf{(b)} Rydberg laser frequency noise, characterized by measured PSD (cyan) with an in-loop PDH error signal. To take into account finite probe time, we smoothen the PSD (dark blue) with a moving average window of 20~kHz, corresponding to 50 $\mu \text{s}$, the upper bound of the spin-locking time used in this specific experiment.
    } 
    \vspace{-0.2cm}
    \label{fig:spin_lock}
\end{figure}
\medskip
\textit{Probing the frequency noise PSD with spin-locking.} --- Beyond its application to two-qubit gates, FRT can be used to explain single-qubit spin-lock~\cite{Bodey2019,Finkelstein2024} experiments, which provide an alternative way of measuring laser frequency noise directly from an atomic observable up to maximally applicable Rabi frequency, complementary to existing methods~\cite{Bishof2013}. In a spin-lock sequence, atoms are initialized to an equal superposition of the Rydberg state and the \clock\ state. We then apply a continuous drive, parallel to the state vector, at a test Rabi frequency $\Omega = 2\pi \times f_{\text{test}}$ before a final $\pi/2$ readout pulse, resulting in a characteristic decay signal due to frequency noise (Appendix~\ref{Appendix:spin_lock}). 

FRT predicts that in the limit of long evolution time, the decay rate of the spin-lock signal is $\pi^2 S_\nu(f_{\text{test}}) + \Gamma_{\text{Ryd}}/2$, where $S_\nu(f_{\text{test}})$ is the frequency PSD at the test Rabi frequency and $\Gamma_{\text{Ryd}}$ is the Rydberg decay rate. We test this prediction by comparing it with the spin-lock signal decay rate calculated by the full error model at several test frequencies (dark green curve and red markers in Fig.~\ref{fig:spin_lock}a). We find good agreement between these two predictions, confirming that the sequence indeed extracts the frequency noise PSD.

We further perform such spin-lock experiments with several test Rabi frequencies and find the measured decay rates (purple markers in Fig.~\ref{fig:spin_lock}a) to be consistent with the error model predictions, probing the strength of laser frequency noise directly with an atomic signal, which was otherwise only measured using a Pound-Drever-Hall error signal~\cite{Madjarov2020,deLeseleuc2018,Shaw2024B}.

\begin{figure}[ht]
    \centering
    \includegraphics[width=\columnwidth]{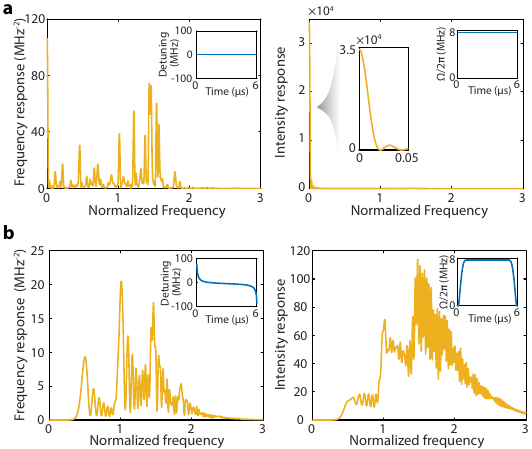}
    \caption{\textbf{Many-body response functions for a Rydberg quantum simulator.} \textbf{(a)} Response functions for quench dynamics of a 7-qubit system. The system is initialized to state $\ket{0}^{\otimes 7}$ and a constant quench Hamiltonian with $7.7$~MHz Rabi frequency is turned on for 6~$\mu$s. The system is most sensitive to slow intensity noise because its effect coherently accumulates with time. \textbf{(b)} Response functions for 6~$\mu$s quasi-adiabatic preparation of a $\mathbb{Z}_2$ state in a 7-qubit system. The detuning sweeps from ${\approx} +10\Omega$ to ${\approx} -10\Omega$. The system is most sensitive to noise above a certain frequency, which is the many-body gap at the phase transition (see text). Insets: Detuning and Rabi frequency sweeps of the simulated dynamics.
    }
    \vspace{-0.2cm}
    \label{fig:MB}
\end{figure}
\medskip
\textit{Many-body dynamics.} --- FRT is not restricted to single- and two-qubit operations and is applicable for predicting the preparation fidelity of many-qubit experiments, for example in the context of analog quantum simulation~\cite{Shaw2024C,Bernien2017}. In this case, qubits are encoded by the \clock\ state $\ket{0_i}$ and Rydberg state ($n^3\mathrm{S}_1$) $\ket{1_i}$ at site $i$. The Hamiltonian describing the system is 
\begin{equation}
    \hat{H}= \dfrac{\Omega}{2}\sum_i \hat{\sigma}_i^x - \Delta\sum_i \hat{n}_i + \sum_{i < j} \dfrac{C_6}{r_{ij}^6} \hat{n}_i \hat{n}_j
\end{equation}
where $ \hat{\sigma}_i^x = (\ket{0_i}\bra{1_i} + \ket{1_i}\bra{0_i}) $ and $\hat{n}_i = \ket{1_i}\bra{1_i}$.

Here, we consider and compare two types of many-body dynamics: quench~\cite{Shaw2024C} and adiabatic $\mathbb{Z}_2$ state preparation~\cite{Bernien2017} with a detuning sweep in a 7-qubit array. For both cases, the system is initialized in a product state $\ket{0}^{\otimes 7}$. For the quench dynamics, a Rydberg Rabi frequency of $7.7~\text{MHz}$ and zero detuning is turned on suddenly at $t=0$ and lasts for $6~\mu\text{s}$. For the adiabatic $\mathbb{Z}_2$ state preparation, a tangent detuning sweep from $+10\Omega$ to $-10\Omega$ is applied over $6~\mu\text{s}$.

Response functions for these two dynamics (Fig.~\ref{fig:MB}) exhibit distinctive features. For the quench dynamics, slow intensity noise builds up coherently over time, leading to a strong response at low frequency. For the adiabatic sweep, since the state approximately follows the ground state of the Hamiltonian, it is sensitive to noise which is resonant with energy differences between the instantaneous ground state and excited eigenstates. The minimal excitation frequency to the lowest excited state appears at the phase transition from disordered to $\mathbb{Z}_2$ phase. Hence, there exists a cutoff frequency, corresponding to this many-body gap at the phase transition, below which the sequence is insensitive to noise.

\section{Towards 0.999 CZ fidelity}\label{Sec:to999}
\begin{figure}[ht!]
    \centering
    \includegraphics[width=\columnwidth]{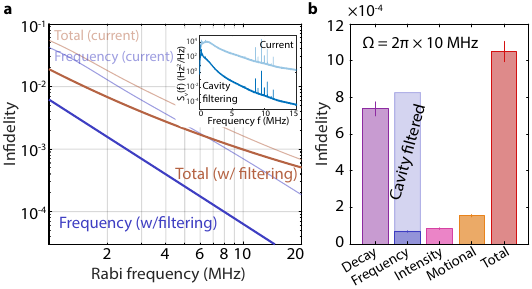}
    \caption{\textbf{Rydberg laser frequency noise suppression and realistic immediate route to $\mathbf{{\gtrsim} 0.999}$ fidelity for CZ gates.} 
    \textbf{(a)} Frequency-noise-induced infidelity and the total infidelity before and after frequency noise suppression via cavity filtering. All four curves are obtained from FRT with an \textit{ideal} time-optimal gate. With all other error sources fixed, this indicates that ${\gtrsim}0.999$ fidelity with slightly higher Rabi frequencies, for example 10~MHz, is within reach. Inset: Current Rydberg laser frequency PSD and projected PSD filtered by a cavity with 140~kHz linewidth.
    \textbf{(b)} Simulated error budget (from a full error model) for a \textit{realistic} CZ gate operating at 10~MHz with laser frequency noise suppression via cavity filtering. The full error model predicts the effects of frequency noise suppression as well as infidelity contributions from other major error sources. It shows that a gate fidelity ${\gtrsim} 0.999$ is readily reachable.
    }
    \vspace{-0.2cm}
    \label{fig:cavity_filter}
\end{figure}

Finally, with our \textit{ab initio} error model and FRT, we propose a pathway for achieving an entangling gate fidelity ${\gtrsim}0.999$ in near-term neutral atom array platforms. Since infidelity due to Rydberg decay scales approximately as $\Omega^{-1}$ and infidelity due to frequency noise and finite atomic temperature scales approximately as $\Omega^{-2}$, a higher Rabi frequency is beneficial. Reducing laser frequency PSD is also readily achievable by using a fiber laser or performing cavity filtering~\cite{Levine2019}. The FRT calculation predicts that by applying cavity filtering to our current laser and increasing the Rabi frequency to 10~MHz, we can achieve fidelity of ${\gtrsim} 0.999$ (Fig.~\ref{fig:cavity_filter}a). A full \textit{ab initio} error model simulation with finite rise time and finite blockade interaction further confirms this prediction (Fig.~\ref{fig:cavity_filter}b), showing a realistic route to useful applications requiring lower qubit error rates.

\section{Summary and Outlook}\label{Sec:outlook}
In this work, we have demonstrated a new state-of-the-art CZ gate fidelity of \highFidBBCorrected, averaged over the two-qubit symmetric (with respect to exchange of the two qubits) subspace, on a neutral atom array platform. With our numerical \textit{ab initio} error model~\cite{Shaw2024B}, we traced down and studied relevant error sources, matching experimental observations within error bars. We further tested the presented benchmarking sequence with the same error model by comparing it with other standard metrics for the CZ gate fidelity.

The ability to accurately benchmark gate fidelity and match error model predictions across a range of gate speeds further enables us to project a realistic route to achieving a two-qubit gate fidelity of ${\gtrsim}0.999$. We find that this can be achieved with readily available upgrades to our laser system, by reducing frequency noise~\cite{Li_Covey2022,Levine2019} and moderately increasing Rabi frequency.

We have shown that the symmetric stabilizer benchmarking sequence presented in this work yields a good proxy of the CZ fidelity averaged over the symmetric Haar random subspace for our experiment-specific error mechanisms. Generalizing to generic error models may require circuits with Clifford randomization~\cite{Magesan2012,Gaebler2012}, under which the conversion of error channels to depolarizing noise is better than the case of generators. Overcoming sensitivity to errors in the randomization gates, in that case, may be achieved by applying a similar principle in circuit design as we have employed here. For instance, one can replace $\pi/2$ pulses with global single-qubit Clifford gates for better randomization. Furthermore, to benchmark the CZ fidelity over the full two-qubit Hilbert space, one can design an initialization unitary to draw a random two-qubit stabilizer state~\cite{Gottesman1997} with a uniform probability distribution and use local single-qubit Clifford or two-qubit Clifford gates for randomization.

We have further developed an analytical fidelity response theory (FRT) to predict the effect of laser noise on gate fidelity in a computationally resource-efficient manner. Treating it as a toolbox, one may simply take the specific laser noise PSD on their experiment and scale the approximate functional forms of the response functions accordingly (provided in Appendix~\ref{Appendix:ApproxForms}) to predict how time-optimal gates~\cite{Jandura2022,Evered2023} would be affected by laser noise. This efficiency may also enable combining the toolbox with optimal control techniques~\cite{Muller2022, Caneva2011} by recasting response functions as cost functions, along with specific laser noise PSD, to design experiment-specific gate protocols robust to laser noise. Furthermore, our framework can be readily extended to complement existing optimization methods on other platforms such as trapped ions~\cite{Kang2023} and superconducting qubits~\cite{Hyyppa2024}, for noise sources that can be characterized by a PSD~\cite{Nakav2023,Day2022}.

Further, we expect measuring frequency noise with a spin-lock sequence --- as demonstrated here --- to become a standard technique, eliminating uncertainties in the corresponding PSD, which is typically only known from indirect measurements~\cite{deLeseleuc2018,Shaw2024B}. The developed FRT could also see wider range of applications in the context of studying quantum many-body physics problems with quantum simulators, for which we demonstrated a first numerical proof-of-principle here. 

Our results could be widely applicable to benchmark and predict quantum device performance, uncover dominant error sources and their effects, and guide the design of future experiments toward even higher-fidelity operations. \\

\section*{Acknowledgements}
We thank Kon Leung, Dariel Mok, Gyohei Nomura, Ingo Roth, Jeff Thompson, and Johannes Zeiher for fruitful discussions and their feedback on this work. We acknowledge support from the DARPA ONISQ program (W911NF2010021), the DOE (DE-SC0021951), the Army Research Office MURI program (W911NF2010136), the NSF QLCI program (2016245), the Institute for Quantum Information and Matter, an NSF Physics Frontiers Center (NSF Grant PHY-1733907), and the Technology Innovation Institute (TII). Support is also acknowledged from the U.S. Department of Energy, Office of Science, National Quantum Information Science Research Centers, Quantum Systems Accelerator. RBST acknowledges support from the Taiwan-Caltech Fellowship. RF acknowledges support from the Troesh postdoctoral fellowship.


\newpage
\setcounter{section}{0}
\setcounter{equation}{0}
 
\captionsetup[figure]{labelfont={bf},name={FIG.},labelsep=period,justification=raggedright,font=small}
\renewcommand{\thesection}{Appendix \Alph{section}}
\renewcommand{\thesubsection}{\arabic{subsection}}
\renewcommand{\theequation}{A\arabic{equation}}

\newpage
\clearpage
\appendix
\section{Experimental setup and data analysis}\label{Appendix:Setup}
Our experimental setup has been detailed in previous works~\cite{Scholl2023A, Shaw2024, Finkelstein2024}. In particular, the qubit considered here is defined on the optical clock transition ($^1\text{S}_0$ $\leftrightarrow$ $^3\text{P}_0$) (Fig.~\ref{fig:overview}d). The laser, addressing this transition and performing single-qubit rotations, is thus commonly referred to as the \textit{clock laser} in this work. The clock transition Rabi frequency applied in the CZ gate benchmarking sequence is about 2.1~kHz. Relevant to the entangling gate benchmarking sequence in this work, global single-qubit rotations in the set $\{R_{\pm X}(\pi/2), R_{\pm Y}(\pi/2)\}$ were benchmarked in our previous work~\cite{Finkelstein2024}. Applying a pulse train of randomly chosen gates from this set, we previously obtained a fidelity of 0.9978(4), comparable to the entangling gate fidelity, as emphasized in Section~\ref{Sec:benchmarking}.

During the CZ gate, we switch off the tweezer trap temporarily with an AOM and a rise time on the order of 50 ns. This is to eliminate additional light shift induced by tweezers. For error modeling during the CZ gates, we thus consider only the atomic motion with finite temperature and tweezers switched off in the scope of this work.

\begin{figure*}[ht!]
    \centering
    \includegraphics{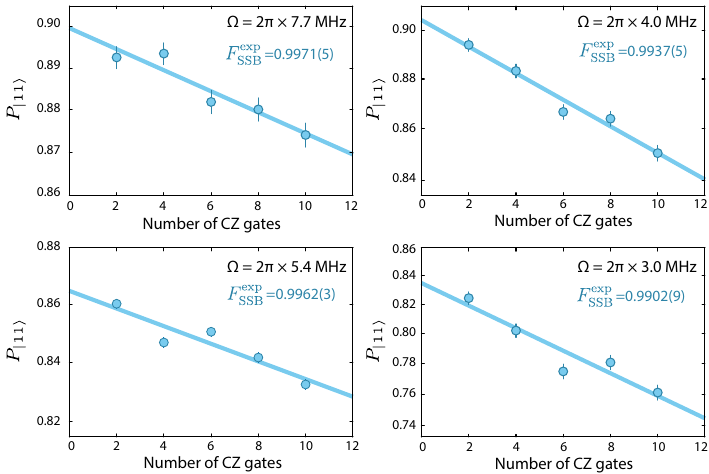}
    \caption{\textbf{Experimental data for symmetric stabilizer benchmarking sequence with CZ gates at different Rydberg Rabi frequencies.} For each Rydberg Rabi frequency applied, we benchmark the time-optimal CZ gate with the SSB sequence. We use the maximum-likelihood method (see text) to fit an exponential decay to $P_{\ket{11}}$ as a function of $N_\mathrm{CZ}$ and obtain a fitted decay with a standard error. After correcting for false contribution from leakage during the gate (see text and Appendix~\ref{Appendix:BrightState}), we obtain $F_\mathrm{SSB}^\mathrm{exp}$ and the associated standard error, representing one-sigma confidence intervals. We further provide in the text the reduced chi-squares $\chi^2_\mathrm{red}$ for each fit.}
    \vspace{-0.1cm}
    \label{SI_SSB_raw_data}
\end{figure*}

For each experimental data point at a given Rydberg Rabi frequency presented in Fig.~\ref{fig:overview}f, we optimize the CZ gate parameters first with an echo benchmarking sequence~\cite{Finkelstein2024}. Then, we calibrate the single-atom phase directly on the SSB sequence (Appendix~\ref{Appendix:SSB}). Once we have a gate calibrated, we perform data-taking with a varying number of CZ gates applied (Fig.~\ref{SI_SSB_raw_data}), interleaved with feedback runs~\cite{Finkelstein2024,Choi2023}, including clock resonance frequency calibration and Rydberg resonance frequency calibration. The total number of applied CZ gates, $N_\mathrm{CZ}$, includes the two CZ gates in $\hat{U}_\mathrm{init}$ and $\hat{U}_\mathrm{rec}$ (Appendix~\ref{Appendix:SSB}).

To extract $F_\mathrm{SSB}^\mathrm{exp}$ from the experimental data, we fit it using the maximum-likelihood method~\cite{Scholl2023A, Finkelstein2024} to the model function $P_{\ket{11}} = a_0\times F^{N_\mathrm{CZ}}$, for $N_\mathrm{CZ}\geq2$. (Note that a model function $P_{\ket{11}} = b_0\times F^{N_\mathrm{CZ}-2}$ yields identical results for $F$.) The error bar obtained from this method represents one-sigma confidence intervals. For completeness, here we further provide the reduced chi-squared test result $\chi^2_\mathrm{red}$ for each fit: 1.26, 5.6, 1.32, and 4.07 for the data set at Rydberg Rabi frequency of 7.7~MHz, 5.4~MHz, 4.0~MHz, and 3.0~MHz, respectively.

Finally, we correct for false contribution from leakage during the gate (Appendix~\ref{Appendix:BrightState} and Table~\ref{tab:false_contribution}) to obtain $F_\mathrm{SSB}^\mathrm{exp}$ (Fig.~\ref{SI_SSB_raw_data}).

\begin{figure}[ht!]
    \centering
    \includegraphics[width=\columnwidth]{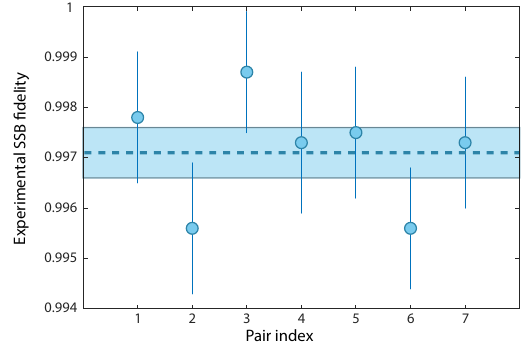}
    \caption{\textbf{Pair-by-pair SSB-extracted fidelity at Rabi frequency \highRabi.} The extracted CZ fidelity for each pair across the one-dimensional array is plotted with the array-averaged fidelity (dashed blue line) and the associated standard error (blue shaded area), obtained from fitting the array-averaged data set.  We further provide in the text the reduced chi-squares $\chi^2_\mathrm{red}$ for each fit.}
    \vspace{-0.1cm}
    \label{SI_SSB_pair_by_pair}
\end{figure}

We also perform a pair-by-pair analysis at the highest Rabi frequency applied 7.7~MHz (Fig.~\ref{SI_SSB_pair_by_pair}), which shows the uniformity of the gate fidelity across the array. We further provide the reduced chi-squared test result $\chi^2_\mathrm{red}$ for each fit: 1.80, 0.40, 0.12, 0.58, 0.38, 0.53, 0.15 for pairs 1 to 7, respectively.

\section{\textit{Ab initio} Rydberg error model}\label{Appendix:Ryd}
Throughout this work, all numerical simulation results (except those from FRT) are performed with our \textit{ab initio} error model, tested extensively in our previous works~\cite{Choi2023,Scholl2023A,Shaw2024C,Finkelstein2024}. A complete description of the model and various error sources is detailed in Chapter 2 of Ref.~\cite{Shaw2024B}. Here, we will summarize the general framework and describe the most relevant error sources, contributing to the CZ gate infidelity.

The error model is based on a Monte Carlo wavefunction approach~\cite{Molmer1993} where we simulate quantum evolution of the system over a large number of randomly generated noise trajectories. For each trajectory, we sample laser noise traces from the power spectral densities (PSD). We then perform a discrete time evolution based on the resulting noisy Hamiltonian. At each time step, we introduce probabilistic quantum jumps to account for spontaneous decays. In this work, all numerical results for \textit{realistic} gates are performed with the Hamiltonian in Eq.~\ref{eq:gate_Ham_realistic}. We typically generate $\gtrsim 500,000$ trajectories for single-CZ-gate simulation and $\gtrsim 10,000$ trajectories for each underlying data point in the benchmarking circuit simulation. The error bars in all presented numerical simulation results are statistical uncertainties of state fidelities (for single gate) or fitting parameter uncertainties (for circuit) with one-sigma confidence intervals.

\medskip
\textit{Rydberg laser intensity noise} --- The laser intensity noise translates to fluctuations in the Rydberg Rabi frequency. In our numerical simulation, we generate a time-dependent Rabi frequency trace from the measured relative intensity noise (RIN) PSD of the Rydberg laser in use (Fig.~\ref{fig:linear_response}b, lower). Additionally, we add a random offset to the Rabi frequency, following a Gaussian distribution of standard deviation 0.8\%, which accounts for the DC low-frequency (shot-to-shot) intensity noise. This value and the RIN PSD are measured with a Si-amplified photodiode (Thorlabs PDA8A2) placed as close to the atoms as possible.

\medskip
\textit{Rydberg laser frequency noise} --- The laser frequency noise translates directly to fluctuating detuning. We use a PSD for the quantitative description of laser frequency noise. To obtain this, we measure the Pound-Drever-Hall (PDH) locking error signal, referenced to an IR cavity~\cite{Madjarov2020} with a pick-off from the seed laser at 1267 nm. We then perform Fourier analysis, take into account cavity roll-off factors of the reference cavity and the two following doubling cavities, and average over hundreds of traces to obtain the PSD (Fig.~\ref{fig:linear_response}b, upper). DC low-frequency frequency noise, as shown in Fig.~\ref{fig:linear_response}c, is measured with an atomic signal~\cite{Shaw2024B}. Finally, we sample a time-dependent detuning noise trace from this PSD~\cite{deLeseleuc2018} and the DC part for the noisy trajectory simulation. 

Since there are generally several electronic and optical components between the laser and the atoms, one might reasonably suspect that the atoms experience a different PSD. In addition, electric field noise could add to detuning noise. An experimental way of testing the frequency noise PSD is the spin-lock experiment. As an application of FRT developed in this work, we compare the analytical prediction, full numerical simulation results, and the experimental results of the Rydberg spin-lock experiment (Fig.~\ref{fig:spin_lock}). Details on the experimental sequence and the analytic prediction are given in Appendix~\ref{Appendix:spin_lock}.

\medskip
\textit{Rydberg spontaneous and blackbody-radiation-induced decay} --- We simulate the decay using the quantum jump method. For each time interval $dt$, the decay channel $\ket{i} \rightarrow \ket{f}$ occurs with probability $\Gamma_{if} |\braket{i}{\psi(t)}|^2 dt$, where $\ket{\psi(t)}$ is the quantum state at time $t$. If a quantum jump happens, we will apply the projection operator $\ketbra{f}$ to the state. The following quantum jumps are most relevant and included in the simulation. These numbers are all measured with atomic signals as detailed in our previous work~\cite{Scholl2023A} with the same principal quantum number $n=61$: \textit{i)} Rydberg to dark state decay, with lifetime $\tau = 78\mu s$; here, dark state refers to states which are dark under fluorescence imaging, such as nearby Rydberg states; \textit{ii)} Rydberg to $^3\text{P}_J$ manifold (bright state) decay, with lifetime $\tau = 166\mu s$ and branching ratio $10\%,\ 30\%,\ 60\%$ for $J=0, 1, 2$; here, bright state refers to states that are bright during fluorescence imaging; \textit{iii)} $^3\text{P}_1$ to ground state $^1\text{S}_0$ decay, with lifetime $\tau = 21\mu s$; \textit{iv)} $^3\text{P}_1$ and $^3\text{P}_2$ to dark state ionization, with $\tau = 320\mu s/(\Omega/\mathrm{MHz})^2$; here, dark state ionization refers to excitation that leads to ionization and makes the atom dark under fluorescence imaging. 

\medskip
\textit{Atomic motion} --- For atomic motion, we assume a time-of-flight dynamics after turning off the trap when a CZ gate is applied. We sample the displacement and the velocity of the atoms accordingly. The velocity along the Rydberg beam optical axis translates to shot-to-shot detuning fluctuations due to the Doppler effect. Doppler noise is inferred from measured atomic temperature~\cite{Finkelstein2024}; while it appears in the atomic signal used to determine the DC part of frequency noise~\cite{Shaw2024B}, we subtract the Doppler noise contribution from this signal, such that it is \textit{not} included in the DC contribution to frequency noise shown in Fig.~\ref{fig:linear_response}c and the noisy trajectories for frequency noise. The displacement of the atom from the center of the beam (with a finite waist radius of ${\approx} 20$ $\mu$m) translates to shot-to-shot Rabi frequency fluctuations due to beam sampling. 

\section{Single-qubit (clock) gate error model}\label{Appendix:Clock}
We detail here how we include the single-qubit gate errors when we perform full error model simulation on the various benchmarking sequences (Fig.~\ref{SI_corr_clock}, Fig.~\ref{SI_Harvard}). These single-qubit gate errors stem mainly from the frequency noise of the clock laser, used to perform single-qubit rotations. We independently measure the clock laser frequency noise PSD from atomic signals, as detailed in the Methods section of our previous work~\cite{Finkelstein2024}. In particular, we use the Ramsey sequence to characterize the low-frequency noise and the spin-lock sequence to characterize the fast-frequency noise.

In addition to the laser frequency noise, since we operate in a sideband resolved regime, the single-qubit operations are usually sensitive to a finite temperature. However, we perform erasure-cooling~\cite{Scholl2023B} to prepare atoms close to their motional ground state ($\Bar{n}\approx 0.01$). Although this finite temperature leads to a negligible impact on the single-qubit rotation fidelity, we still include it in the clock error model.

Same as in the Rydberg error model, we utilize the Monte Carlo wavefunction approach~\cite{Molmer1993} and generate frequency noise traces to simulate noisy trajectories on the quantum system evolution. For the simulation of single-qubit gates interleaved with CZ gates in the benchmarking sequences, we simulate the three-level system dynamics with these generated traces and the Rydberg laser noise traces.

\section{Comparison of CZ gate fidelity metrics $F_\mathrm{Haar}$ and $F_\mathrm{Sym}$}\label{Appendix:CompareCZ}
As we see in Fig.~\ref{fig:comparison_benchmark}, the CZ gate fidelity averaged over symmetric Haar subspace $F_\mathrm{Sym}$ is different from that averaged over two-qubit Haar random states $F_\mathrm{Haar}$. The difference can be understood from their different Rydberg decay probabilities, different responses to laser noise, and shot-to-shot detuning due to atomic motion. 

\medskip
\textit{Rydberg decay} --- First, we analyze the Rydberg decay contribution to $F_\mathrm{Haar}$ and $F_\mathrm{Sym}$. The decay probability is computed by integrating the population in the Rydberg state over time, multiplied by the Rydberg decay rate, during the course of the time-optimal gate. For an \textit{ideal} gate with zero rise/fall time and infinite blockade energy, we find the decay probability for an initial state of $\ket{01}$ ($\ket{10}$) and that of $\ket{11}$ being $\varepsilon_1 = 3.9\Gamma/\Omega$ and $\varepsilon_2 = 3.9 \Gamma/\Omega$, respectively, where $\Gamma$ is the Rydberg state decay rate. Therefore, the average decay probability for $F_\mathrm{Haar}$ is $(2\varepsilon_1 + \varepsilon_2)/4 = 2.9 \Gamma/\Omega$ while that for $F_\mathrm{Sym}$ is $(\varepsilon_1 + \varepsilon_2)/3 = 2.6 \Gamma/\Omega$. For the decay contribution to $F_\mathrm{Haar}$ and $F_\mathrm{Sym}$ to be equal, $2\varepsilon_1=\varepsilon_2$ has to be satisfied, which is not the case for the different conditions studied throughout this work.

\begin{figure}[ht!]
    \centering
    \includegraphics[width=\columnwidth]{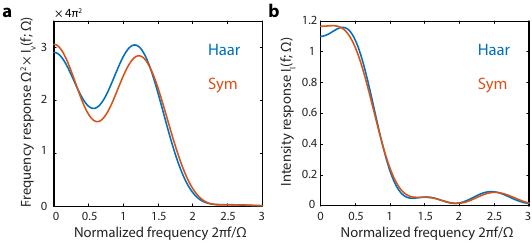}
    \caption{\textbf{Universal response functions of an \textit{ideal} time-optimal CZ gate, averaged over two-qubit Haar random states (blue, Haar) and symmetric Haar random states (orange, Sym).} Both \textbf{(a)} frequency noise response and \textbf{(b)} intensity noise response are presented here. Qualitatively, response functions show similar structures and diminishing trends toward noise frequencies higher than the Rabi frequency (normalized frequency greater than 1). Approximate fitted analytical forms are given in Appendix~\ref{Appendix:ApproxForms}.} 
    \vspace{-0.3cm}
    \label{SI_response_compare}
\end{figure}

\medskip
\textit{Rydberg laser noise} --- The difference in infidelity due to laser noise can be understood from FRT. In Fig.~\ref{SI_response_compare}, we show the frequency and intensity response functions for both Haar random states and symmetric Haar states. Response functions for these two sets of states are slightly different. In particular, the maximum relative frequency response difference within $2\pi f /\Omega \in [0, 1.5]$ is 17\%. In practice, the frequency-noise-induced infidelities for these two sets of states differ by no more than $7\%$ under integration with our frequency PSD. 

\medskip
\textit{Atomic motion} --- The sensitivity to atomic motion, which is dominated by the single-atom shot-to-shot detuning noise due to the Doppler effect, is characterized by the response function of a single-atom detuning at zero frequency. Hence, the effect of atomic motion scales as $\Omega^{-2}$. The response to single-atom detuning for both $F_\mathrm{Haar}$ and $F_\mathrm{Sym}$ is ${\approx} 3.0/[\Omega/(2\pi)]^2$.

\section{Symmetric stabilizer benchmarking}\label{Appendix:SSB}
\renewcommand{\theequation}{E\arabic{equation}}

We provide comprehensive details on the symmetric stabilizer benchmarking (SSB) circuit implementation, including the choices of initialization and recovery unitaries, and on the numerical simulation of circuit-inferred infidelity and symmetric infidelity using both our single-qubit gate error model and two-qubit gate error model. 

The goal of designing a benchmarking circuit for CZ gates is to faithfully extract the Rydberg gate error and to fully isolate the interleaved single-qubit gate errors.
However, there is not yet any known model-free theoretical guarantee that the SSB sequence, as well as other randomized-benchmarking-type circuits ~\cite{Knill2007,Magesan2011,Magesan2012,Baldwin2020,Evered2023,Ma2023}, could extract the CZ gate fidelity in the presence of complex noise power spectral densities  for both single-qubit gates and two-qubit gates and complex decay channels. 
Yet, in Section~\ref{Sec:benchmarking}, we already found that the SSB prediction and the CZ gate fidelity averaged over symmetric Haar states are consistent across a wide Rabi frequency range, in the absence of single-qubit gate errors (Fig.~\ref{fig:comparison_benchmark}) and in the presence of single-qubit gate errors (Fig.~\ref{SI_corr_clock}). As a continuation, we detail our sequence and study the Evered 2023 sequence through numerical simulation with our Rydberg error model and single-qubit gate error model and explain why Evered 2023 is sensitive to single-qubit gate error while our circuit has reduced sensitivity with simple, analytical error models.

The section is structured as follows: first, we explain the details of the SSB sequence implementation; next, we simulate the Evered 2023 sequence~\cite{Evered2023} with our error model and find their sequence sensitive to single-qubit gate errors. We then propose an improvement to minimize its sensitivity to single-qubit gate errors, which enables direct comparison between the experimental results. Further, we analyze a few analytical single-qubit gate and two-qubit gate error models to illustrate why Evered 2023 is sensitive to single-qubit gate errors, and how our sequence is designed to reduce sensitivity to single-qubit gate errors. We find that our reduced sensitivity can be explained by the constant distribution of state throughout the sequence. Finally, we numerically study the effect of temporally coherent noise between one and subsequent CZ gates in one single instance of the benchmarking circuit.

\subsection{Details of the SSB circuit}
In this subsection, we explain the details of the SSB sequence, as well as the initialization and recovery unitary in the circuit (Fig.~\ref{SI_SSB_seq}a). There are in total twelve symmetric stabilizer states (SSS) as per our definition, all of which are listed in Table~\ref{tab:SSS}.

The unitary $\hat{U}_\mathrm{init}$ ($\hat{U}_\mathrm{rec}$) is composed of two single-qubit $\pi/2$ rotations, followed by a CZ gate, and then three (two) single-qubit $\pi/2$ rotations. The rotations chosen are listed in Table~\ref{tab:Uinit},~\ref{tab:Urec}.

\begin{table}[ht!]
    \centering
    \begin{tabular}{c|ccccc}
    \hline
       Stabilizers  &  $R_1$ & $R_2$ & $R_3$ & $R_4$ & $R_5$ \\
       \hline
    $IX, XI$ & X & X & Y & -Y & Y\\
    $-IX, -XI$ & X & -X & Y & -Y & Y\\
    $IY, YI$  & X & -X & X & -X & X\\
    $-IY, -YI$ & X & X & X & -X & X\\
    $IZ, ZI$ & X & X & X & -Y & -X\\
    $-IZ, -ZI$ & X & -X & X & -Y & -X\\
    $XZ, ZX$ & -X & -Y & X & -Y & -X\\
    $-XZ, -ZX$ & X & Y & X & -Y & -X\\
    $YZ, ZY$ & -X & -Y & X & -X & X\\
    $-YZ, -ZY$ & X & Y & X & -X & X\\
    $XY, YX$ & X & Y & Y & -Y & Y\\
    $-XY, -YX$ & -X & -Y & Y & -Y & Y\\
    \hline
    \end{tabular}
    \caption{\textbf{Rotations chosen for $\hat{U}_\mathrm{init}$}. For $R_i$ in this table, X, -X, Y, and -Y represent the shorthand notations for $R_{X}(\pi/2)$, $R_{-X}(\pi/2)$, $R_{Y}(\pi/2)$, and $R_{-Y}(\pi/2)$, respectively.}
    \vspace{-0.2cm}
    \label{tab:Uinit}
\end{table}

\begin{table}[ht!]
    \centering
    \begin{tabular}{c|cccc}
    \hline
    Stabilizers  & $R_1$ & $R_2$ & $R_3$ & $R_4$ \\
    \hline
    $IX, XI$ & X & -Y & X & X \\
    $-IX, -XI$ & X & Y & X & X \\
    $IY, YI$  & Y & X & X & X \\
    $-IY, -YI$ & Y & -X & X & X \\
    $IZ, ZI$ & -X & X & X & X \\
    $-IZ, -ZI$ & X & X & X & X \\
    $XZ, ZX$ & -X & Y & Y & X \\
    $-XZ, -ZX$ & X & Y & Y & X \\
    $YZ, ZY$ & Y & Y & Y & X \\
    $-YZ, -ZY$ & X & X & Y & X \\
    $XY, YX$ & -Y & X & Y & X \\
    $-XY, -YX$ & Y & X & Y & X \\
    \hline
    \end{tabular}
    \caption{\textbf{Rotations chosen for $\hat{U}_\mathrm{rec}$}. For $R_r$ in this table, X, -X, Y, and -Y represent the shorthand notations for $R_{X}(\pi/2)$, $R_{-X}(\pi/2)$, $R_{Y}(\pi/2)$, and $R_{-Y}(\pi/2)$, respectively.}
    \vspace{-0.5cm}
    \label{tab:Urec}
\end{table}

In the experiment, each $\hat{U}_\mathrm{init}$ is drawn from the twelve pulse strings with equal probability and each subsequent rotation is drawn from four $\pi/2$ pulses $\{R_{\pm X}(\pi/2), R_{\pm Y}(\pi/2)\}$ with equal probability. $\hat{U}_\mathrm{rec}$ is then calculated depending on the stabilizer state before it. The number of applied CZ gates, $N_\mathrm{CZ}$, includes those in $\hat{U}_\mathrm{init}$ and $\hat{U}_\mathrm{rec}$, thus starting at two (Fig.~\ref{fig:overview}b). With this setup, in the absence of errors, the distribution of states at any stage between $\hat{U}_\mathrm{init}$ and $\hat{U}_\mathrm{rec}$ is a uniform distribution over the twelve SSS.

The Rydberg pulse leaves a single-atom phase $\phi$ on the optical qubit manifold in addition to the well-defined CZ gate~\cite{Levine2019} (Fig.~\ref{SI_SSB_seq}a). A CZ gate is thus composed of a Rydberg time-optimal CZ pulse and a virtual $Z$-rotation, imparted on the clock laser phase, to compensate for this Rydberg-induced single-atom phase. By adding $-\phi$ to the clock laser phase, the rotation axis of subsequent single-qubit rotations is rotated by $-\phi$ in the $x-y$ plane. In the circuit diagrams presented in this work, these virtual $Z(\phi)$ are shown separately from the standard CZ gate notation to take into account these experimental actions. We experimentally scan the applied phase $\phi$ in these virtual gates and calibrate the value that maximizes the return probability (Fig.~\ref{SI_SSB_seq}b).

Calibration errors on this virtual gate are regarded as errors on the implemented CZ gate. In practice, this phase is calibrated precisely, up to $10$ mrad on the experiment, corresponding to $4\cdot10^{-5}$ infidelity on the CZ gate. Experimental miscalibration contributes thus marginally to the CZ gate infidelity. This effect is thus not explicitly discussed in the main text.

\begin{table*}[ht!]
\setlength{\tabcolsep}{12pt}
    \centering
    \begin{tabular}{|c|c|c|c|c|c|c|}
    \hline
       $\Omega$ (2$\pi\times$MHz)  & $F_\mathrm{SSB}^\mathrm{exp}$ & $F_\mathrm{SSB}^\mathrm{sim}$ & $F_\mathrm{Evered\ 2023}$ & $F_\mathrm{SSS}$  & $F_\mathrm{Sym}$ & $F_\mathrm{Haar}$ \\
       \hline
       3.0  & 0.9902(9) & 0.99099(17) & 0.99078(17) & 0.99081(7)  & 0.99083(7) & 0.99042(7)   \\
       4.0  & 0.9937(5) & 0.99409(25) & 0.99417(42) & 0.99406(6) & 0.99411(6) & 0.99382(6)   \\
       5.4  & 0.9962(3) & 0.99614(7) & 0.99594(17) & 0.99613(5)  & 0.99602(5) & 0.99598(6)   \\
       7.7  & 0.9971(5) & 0.99751(10) & 0.99746(20) & 0.99750(4) & 0.99754(4) & 0.99742(5)   \\
       \hline
    \end{tabular}\\
    \caption{\textbf{Comparison between experimental results and numerical error model simulation results.} The values shown are the numerical values used in Fig.~\ref{fig:comparison_benchmark}.}
    \label{tab:numbers_comparison}
    \vspace{-0.2cm}
\end{table*}

\subsection{Reduced circuit sensitivity to single-qubit gate errors and comparison with other benchmarking sequences}

To show that our circuit can faithfully predict the CZ gate fidelity, we simulate the circuit prediction of infidelity $1-F_\mathrm{SSB}^\mathrm{sim}$ and compare with the symmetric infidelity $1-F_\mathrm{Sym}$, with only two-qubit Rydberg gate error model (green markers in Fig.~\ref{SI_corr_clock}) and with both two-qubit Rydberg and single-qubit clock gate error models (red markers in Fig.~\ref{SI_corr_clock}). With the full single-qubit gate error model and two-qubit error model, numerical simulation shows that the discrepancy between the circuit-inferred infidelity and the actual gate infidelity, for the lowest fidelity examined (at lowest Rydberg Rabi frequency of 3.0~MHz), is only $5(2)\times 10^{-4}$ ($5(2)\%$ relative difference in infidelity) (Fig.~\ref{SI_corr_clock}). 

\begin{figure}[htbp!]
    \centering
    \includegraphics{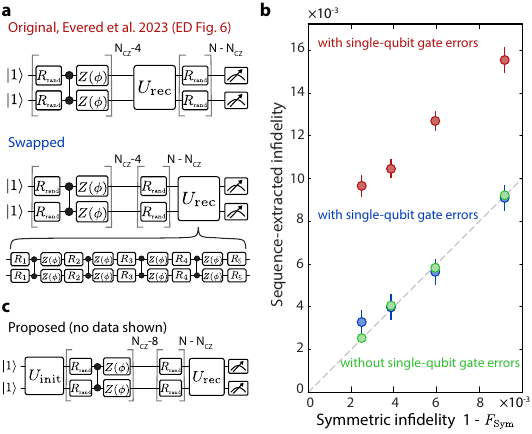}
    \caption{\textbf{Correlation between the extracted fidelity from Evered 2023 sequence and the symmetric fidelity in numerical simulation.}
    \textbf{(a)} The original circuit presented in Ext. Data Fig. 6 of Ref~\cite{Evered2023} and a modified circuit, where the order of the recovery unitary (which contains four CZ gates) and subsequent Haar random single-qubit rotations is swapped.
    \textbf{(b)} We simulate the Evered 2023 benchmarking sequence at four Rabi frequencies used in this work and extract infidelities from fitting the return probability versus the number of applied CZ gates. We then compare these values with infidelities averaged over symmetric Haar states at these Rabi frequencies. Dashed line is a guide to the eye for 1:1 ratio. Without single-qubit gate errors (green), the sequence is a good proxy of $F_\mathrm{Sym}$. With single-qubit gate errors (in our system), the sequence infers higher infidelities (red). By using the swapped circuit, the infidelity is practically insensitive to single-qubit gate errors (blue).
    \textbf{(c)} To further minimize sensitivity to single-qubit gate errors, we propose adding an initialization unitary $\hat{U}_\mathrm{init}$, similar to the structure of $\hat{U}_\mathrm{rec}$, which creates an initial Haar random distribution over the symmetric subspace.
    }
    \vspace{-0.4cm}
    \label{SI_Harvard}
\end{figure}

Such reduced sensitivity to single-qubit gate errors is due to the design that we intentionally keep the distribution over states unchanged in our circuit (detailed in the next subsection) and is therefore not a feature for a few other circuits.
For instance, as shown in Fig.~\ref{fig:comparison_benchmark}, we simulate the Evered 2023 sequence, presented in Ext. Data Fig. 6 of Ref~\cite{Evered2023}. In the absence of single-qubit gate errors and under our Rydberg error model, their circuit also infer fidelities that are consistent with $F_\mathrm{Sym}$ (Fig.~\ref{SI_Harvard}b, green markers).

However, if we were to implement in our experiment the circuit as presented therein, we would obtain higher infidelities due to single-qubit gate errors (Fig.~\ref{SI_Harvard}b, red markers). The sensitivity stems from the fact that the time spent in an entangled state is not fixed but increases as the number of CZ gates increases, which leads to gradually larger effects due to single-qubit gate errors (detailed in the next subsection). This will not be crucial if single-qubit operations are of high fidelity ($>0.9998$ as many hyperfine qubit experiments have reported~\cite{Manetsch2024,Nikolov2023}) but will have a larger impact on our system ($\simeq 0.998$). Therefore, the result presented in this work can be directly comparable with other works as long as they use global benchmarking sequence~\cite{Evered2023,Ma2023} and have high-fidelity single-qubit operations.

We propose a modification to their circuit (\textit{swapped} sequence in Fig.~\ref{SI_Harvard}a), by swapping the $\hat{U}_\mathrm{rec}$ containing four CZs and the subsequent single-qubit rotations. We find that this practically eliminates the sensitivity to single-qubit operations (Fig.~\ref{SI_Harvard}b, blue markers). We speculate the sensitivity could be further reduced by introducing a state preparation unitary to generate an initial state with Haar measure on the symmetric subspace (Fig.~\ref{SI_Harvard}c), the importance of which is detailed in the next subsection. We further note that our SSB sequence, which only utilizes global $\pi/2$ rotations, is easier to implement in our setup, due to fixed single-qubit gate duration.

\subsection{Abstract analytical models for the circuit-inferred infidelity under erroneous single-qubit gates}
In this subsection, we establish a more quantitative understanding of how single-qubit gate errors affect circuit-inferred infidelity. Although the actual error model is complicated, we find a couple of abstract error models for which we can illustrate the reduced sensitivity to single-qubit gate errors under constant distribution of states throughout the circuit.

To start with, we consider a scenario where our SSB circuit (with the return probability fitted to $P_{\ket{11}} = a_0 \times F^{N_{\text{CZ}}}$) yields the exact gate fidelity. Assume that the global single-qubit gate has a two-qubit depolarization channel with depolarizing strength $d_0$, and the Rydberg gate has a leakage channel with probability $\varepsilon$. Then, assuming there are $N$ global single-qubit gates and $N_{\text{CZ}}$ CZ gates, the return probability is
\begin{equation}
    P_{\ket{11}} = \left[ \dfrac{1}{4} + \dfrac{3}{4}(1-d_0 )^N\right] (1-\varepsilon)^{N_{\text{CZ}}}.
\end{equation}
The circuit CZ fidelity inferred from the fit is $F_{\mathrm{SSB}} = 1- \varepsilon = F_{\mathrm{Sym}}$. For the error model discussed above, many benchmarking sequences used in literature \cite{Knill2007, Magesan2011,Magesan2012,Baldwin2020,Evered2023,Ma2023} yield the correct result. 

Next, we find a pedagogical error model which best illustrates why our sequence, in particular, is less sensitive to single-qubit gate errors than some previously experimentally implemented sequences in neutral atom array platforms~\cite{Evered2023,Ma2023}. For a generic single-qubit gate error channel, the fidelity of the final state after the gate depends on the input state (as opposed to a full depolarizing channel). For the predicted fidelity to be insensitive to single-qubit errors, the return probability of the circuit should be insensitive to the number of perfect CZ gates in the sequence. This can be achieved by keeping the input state distribution to all single-qubit gates fixed and invariant when changing the number of CZ gates. 

For instance, consider the error channel of single-qubit gate being a phase-flip error with probability $p$ and two-qubit gates being perfect. For a two-qubit product state $\ket{\psi} \otimes \ket{\psi}$ where $\ket{\psi}$ is sampled over single-qubit Haar measure, the phase-flip channel leads to a fidelity of $f_{\text{prod}} = 1 - \frac{4}{3}p + \frac{8}{15}p^2$. In contrast, for a two-qubit state sampled over symmetric two-qubit subspace with a Haar measure, the phase-flip channel leads to a fidelity of $f_{\text{sym}} =1 - \frac{5}{3}p + p^2$. Hence, for a circuit with $N_1$ global single-qubit gates whose input states are Haar random symmetric two-qubit states and $N - N_1$ global single-qubit gates whose input states are Haar random product states, the fidelity after $N$ global single-qubit gates is approximately
\begin{equation}
    F \approx f_{\text{prod}}^{N-N_1} f_{\text{sym}}^{N_1} + O(p^2) = 1 - \dfrac{4N + N_1}{3}p + O(p^2).
\end{equation}

Hence, for a circuit to claim insensitivity to single-qubit gate errors to the first order in $p$, the circuit needs to keep not only $N$, the total number of single-qubit gates, but also $N_1$ fixed. For the circuit proposed in~\cite{Evered2023} (shown in Fig.~\ref{SI_Harvard}a), this is not the case. Assuming $U_{\text{rec}}$ and CZ gates are perfect, we find $N_1 = N - N_{\text{CZ}}$, which is not a constant. Hence, the circuit inferred infidelity has a constant offset on top of the true infidelity linear in $p$, as can be seen in Fig.~\ref{SI_Harvard}b.

\begin{figure}[htbp!]
    \centering
    \includegraphics{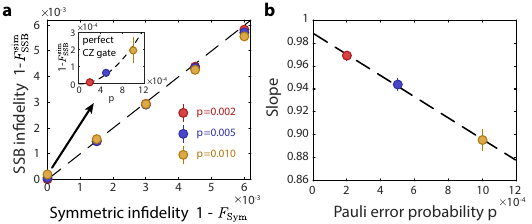}
    \caption{\textbf{Numerical simulation of the abstract analytical model with single-qubit phase-flip error of probability $p$ and two-qubit depolarizing channel.}
    \textbf{(a)} We choose different phase-flip probabilities $p$ and depolarizing channel strengths and simulate the circuit-inferred infidelity $1-F^\mathrm{sim}_\mathrm{SSB}$. The circuit-inferred infidelity is almost linear in the actual infidelity $1-F_\mathrm{Sym} = 0.75d$. Dashed line is a guide to the eye for 1:1 ratio. Inset: the circuit-inferred infidelity for a perfect CZ gate as a function of $p$. Dashed line is a quadratic fit to the data.
    \textbf{(b)} We fit $1-F^\mathrm{sim}_\mathrm{SSB}$ to a linear function in $1-F_\mathrm{Sym}$ while fixing $p$ and extract the slope as a function of $p$. The slope is linear in $p$, with an $y$-intercept slightly smaller than 1, which verifies the prediction Eq.~\ref{eq:SSB_SSS_p}. Dashed line is a linear fit of the data.
    }
    \vspace{-0.1cm}
    \label{SI_phase_flip}
\end{figure}

On the other hand, for the SSB circuit, when assuming $\hat{U}_\mathrm{init}$ and $\hat{U}_\mathrm{rec}$ are perfect, the input states to all single-qubit gates are symmetric two-qubit states. Hence, $N_1 \equiv N$ regardless the number of CZ gates, showing insensitivity to single-qubit gate error up to the linear order.

\begin{table*}[ht!]
\setlength{\tabcolsep}{12pt}
    \centering
    \begin{tabular}{|c|c|c|c|}
    \hline
       $\Omega$ (2$\pi\times$MHz)  & $1-F_\mathrm{Sym}$ ($\times 10^{-4}$) & $1-F_\mathrm{SSB}^\mathrm{sim}$ ($\times 10^{-4}$) & $1-F_\mathrm{SSB,coh}^\mathrm{sim}$  ($\times 10^{-4}$)\\
       \hline
       3.0  & 5.50(27) & 5.16(31) & 3.90(12)  \\
       4.0  & 3.37(16) & 3.28(21) & 2.54(14)  \\
       5.4  & 1.81(8) & 1.88(14) & 1.36(11)  \\
       7.7  & 0.93(4) & 0.97(10) & 0.80(11)  \\
       \hline
    \end{tabular}\\
    
    \begin{tabular}{|c|c|c|c|}
    \hline
       $\Omega$ (2$\pi\times$MHz)  & $1-F_\mathrm{Sym}$ ($\times 10^{-4}$) & $1-F_\mathrm{SSB}^\mathrm{sim}$ ($\times 10^{-4}$) & $1-F_\mathrm{SSB,coh}^\mathrm{sim}$  ($\times 10^{-4}$)\\
       \hline
       3.0  & 0.66(1) & 0.73(7) & 0.90(9)  \\
       4.0  & 0.67(1) & 0.70(2) & 0.98(4)  \\
       5.4  & 0.60(1) & 0.65(8) & 0.92(4)  \\
       7.7  & 0.72(1) & 0.78(8) & 1.29(9)  \\
       \hline
    \end{tabular}\\
    \caption{\textbf{Infidelities due to shot-to-shot laser frequency noise (upper) and laser intensity noise (lower).} We independently measure the shot-to-shot noise. However, we do not know the correlation of the shot-to-shot noise among the gates across a single instance of the sequence. Here, we assume two extreme cases: 1) the intensity/frequency fluctuation for each gate is independent and 2) the values for all gates in an instance are the same, but randomly drawn for each instance. In the table, $1-F_\mathrm{SSB}^\mathrm{sim}$ stands for case 1 and $1-F_\mathrm{SSB,coh}^\mathrm{sim}$ stands for case 2. The table also includes the actual infidelity $1-F_\mathrm{Sym}$ for reference.}
    \label{tab:infidelity DC noise}
    \vspace{-0.2cm}
\end{table*}

Additionally, we now study an analytical two-qubit gate error model in combination with the single qubit gate phase-flip error. Specifically, we consider the two-qubit gate error being a depolarizing channel with strength $d$. The return probability for an SSB circuit with $N$ single-qubit gates and $N_\mathrm{CZ}$ CZ gates is
\begin{equation}
\begin{split}
    P_{\ket{11}} &= \left(1 - \dfrac{5}{3} Np + O(p^2) \right) \left(1 - d \right)^{N_{\mathrm{CZ}}} \\
    &+ \dfrac{1}{4} \left[1 - \left(1 - d \right)^{N_{\mathrm{CZ}}}\right]\\
    &= 1 - \dfrac{5}{3}Np -\left(\dfrac{3}{4}-\dfrac{5}{3}Np \right)d N_\mathrm{CZ} + O(N_{CZ}^2 d^2, p^2)
\end{split}
\end{equation}
By fitting to the model function, the fidelity inferred from the circuit is
\begin{equation}\label{eq:SSB_SSS_p}
\begin{split}
    F_\mathrm{SSB} &= 1 - \dfrac{\frac{3}{4} - \frac{5}{3}Np}{1- \frac{5}{3}Np}d + O(p^2, N_\mathrm{CZ}d^2) \\
    &= 1 - \left(1 - \dfrac{5}{9} Np + O(N_\mathrm{CZ}d) \right) \dfrac{3}{4}d + O(p^2)
\end{split}
\end{equation}
Comparing with the actual gate fidelity $F_\mathrm{Sym} = 1-\frac{3}{4}d$, there are only additional second order terms in $p$ and $d$, \textit{i.e.}, $1-F_{\mathrm{SSB}} = 1-F_{\mathrm{Sym}} + O(p^2, d^2, pd)$, without an additional term in $O(p)$.

For this model, we find our benchmarking circuit tends to overestimate infidelity for small $d$ and underestimate infidelity for large $d$. We numerically simulate this model for $p \in [0.002, 0.01]$ (corresponding to single-qubit gate fidelity $[0.9933, 0.9987]$) and $d \in [0, 0.008]$ (corresponding to two-qubit fidelity $[0.9940, 1]$) (Fig.~\ref{SI_phase_flip}a). We find that when the CZ gate is perfect, the circuit infers an infidelity that is proportional to $p^2$ (Fig.~\ref{SI_phase_flip}a, inset). We also fit a linear model $1-F_\mathrm{SSB} = s(p) (1-F_\mathrm{Sym}) + c$ for each $p$ and find the slope $s(p)$ linear in $p$ (Fig.~\ref{SI_phase_flip}b).

\subsection{Sensitivity to coherent errors}

It is known that in the presence of temporally correlated errors (coherent errors), even standard randomized benchmarking sequence can provide inaccurate results~\cite{Figueroa-Romero2021}. Here, we numerically simulate the effect of coherent errors and show that the effect is below experimental uncertainty. The shot-to-shot laser intensity and frequency fluctuations~\cite{Shaw2024B} are measured with typical experiment repetition rate ($\sim$1~s) and are included in the \textit{ab initio} error model. Since we do not have direct measurement of intensity noise and frequency noise correlation between one and a subsequent CZ gate within the same circuit (which is on the timescale of $\sim$1~ms, much shorter than the experiment repetition rate (1~s) but much longer than the gate duration ($<$1~$\mu$s)), we simply assume that there is a constant intensity and frequency offset that is either: 1) individually drawn from Gaussian distributions, with standard deviation being the measured shot-to-shot standard deviation for each CZ gate or 2) drawn from the same distributions but is applied to all CZ gates in a single instance of circuit realization. We have listed the numerical simulation results, in comparison with the actual gate infidelity due to shot-to-shot noise, in Table~\ref{tab:infidelity DC noise}. Even at the smallest Rydberg Rabi frequency, the deviation is only $1\times 10^{-4}$, much smaller than the experimental error bar. For all other simulation results presented in this work, we always assume that the shot-to-shot noise for each CZ gate is independently drawn (\textit{i.e.}, case 1).

\section{False contribution to $F_\mathrm{SSB}^\mathrm{exp}$ from Rydberg-decay leakage}\label{Appendix:BrightState}
\begin{table*}[ht!]
\setlength{\tabcolsep}{18pt}
    \centering
    \begin{tabular}{|c|c|c|c|c|}
    \hline
       $\Omega$ (2$\pi\times$MHz)  &  $\varepsilon_\mathrm{false}^\mathrm{exp}$ ($\times10^{-4}$)& $\varepsilon_\mathrm{false}^\mathrm{sim}$ ($\times10^{-4}$) & $F_\mathrm{SSB}^\mathrm{raw}$ &  $F_\mathrm{SSB}^\mathrm{exp}$ \\[0.3ex]
       \hline
       3.0  & 3.4(4) & 3.5(1) & 0.9905(9) & 0.9902(9)  \\
       4.0  & \text{not performed} & 2.2(1) & 0.9939(5) & 0.9937(5) \\
       5.4  & 1.8(4) & 1.7(1) & 0.9964(3) & 0.9962(3)  \\
       7.7  & \text{not performed} & 1.3(1) & 0.9972(5)& 0.9971(5)   \\
       \hline
    \end{tabular}\\
    \caption{\textbf{Correcting false contribution from Rydberg-decay leakage during CZ gate.} When performing state-resolved imaging at the end of the benchmarking sequence, atoms that decayed from the Rydberg state into the long-lived $^3\text{P}_2$ state are identified as $\ket{1}$, contributing \textit{falsely} to the return probability $P_{\ket{11}}$ (see text). At different Rabi frequencies, $\varepsilon_\mathrm{false}^\mathrm{exp}$ and $\varepsilon_\mathrm{false}^\mathrm{sim}$ are the false contribution to one single CZ gate fidelity obtained experimentally and numerically, respectively. $F_\mathrm{SSB}^\mathrm{raw}$ is obtained from fitting the raw data (Fig.~\ref{SI_SSB_raw_data}), and $F_\mathrm{SSB}^\mathrm{exp}$ is the \textit{downward}-corrected experimental SSB fidelity value, reported throughout the work.}
    \vspace{-0.2cm}
    \label{tab:false_contribution}
\end{table*}
To measure the return probability $P_{\ket{11}}$ in the benchmarking sequence, which is the probability that both atoms end up in \clock, we perform state-selective imaging. To do this, we apply a push-out beam~\cite{Madjarov2019} to expel all atoms in the ground state \ground, optically pump atoms from \clock, \rstate, and \mstate, and image the remaining atoms. However, spontaneous decay from the Rydberg state populates \rstate\ and \mstate, leading to false contributions to the measurement of $P_{\ket{11}}$. We characterize the false contribution to $P_{\ket{11}}$ and their effect on the extracted CZ fidelity.

Since \rstate\ has a relatively short lifetime (21$\mu$s) and decays to \ground, it does not contribute to these false contributions. However, \mstate\ is a long-lived state and its population accumulates linearly after each CZ gate and contributes falsely in the final image. We experimentally, in the SSB sequence, measure the population in \mstate\ by applying multiple rounds of clock $\pi$-pulse and \ground\ push-out. This depletes the atoms in the ground-clock manifold. The remaining population bright to imaging beam is then in \mstate\ only. We perform the SSB sequence with a different number of CZ gates and measure the \mstate\ population accumulation $\varepsilon_\mathrm{image}$ per pair, per gate.

The false return probability contribution, to the first order, is $P_{\ket{1, ^3\text{P}_2}} + P_{\ket{^3\text{P}_2, 1}}$. However, from the above method of imaging \mstate, we can only infer $P_{\ket{1, ^3\text{P}_2}} + P_{\ket{^3\text{P}_2, 1}} + P_{\ket{0, ^3\text{P}_2}} + P_{\ket{^3\text{P}_2, 0}}$  Conditioned on one atom decaying into the \mstate, the state that the other atom ends up in is random, so it has $1/2$ probability to end up in $\ket{1}$. Therefore, the false contribution to fidelity is $\varepsilon_\mathrm{false}=\frac{1}{2}\varepsilon_\mathrm{image}$.

We list the measurement of false contribution for two different Rabi frequencies and numerical simulation of false contribution to the CZ fidelity for all Rabi frequencies used in this work (Table~\ref{tab:false_contribution}), which are consistent with each other. We then down-correct all measured fidelities with the predictions from the error model. The corrected fidelity values (which are reported throughout the paper everywhere else) and raw-fitted values are also listed in Table~\ref{tab:false_contribution}.

\section{Analytic derivation of fidelity response theory (FRT)}\label{Appendix:LinearResponse}
\renewcommand{\theequation}{G\arabic{equation}}
We derive the response function using first-order perturbation theory (Eq.~\ref{eq:response_func}). We model the noise as a perturbation to the Hamiltonian such that we denote the ideal Hamiltonian as $\hat{H}_0$ and noise as an additional term with amplitude $h(t)$ multiplied by a noise operator $\hat{O}(t)$.
\begin{equation}
    \hat{H}(t) = \hat{H}_0(t) + \delta \hat{H}(t) = \hat{H}_0(t) + h(t) \hat{O}(t)
\end{equation}
Here, the noise operator $\hat{O}(t)$ only depends on the type of noise but not the amplitude of noise and is Hermitian, as implied by the unitary nature mentioned in the main text.

Let $\ket{\psi(t)}$ be the system wavefunction. We move into the interaction picture by taking $\ket{\psi} = \hat{U}_0(t) \ket*{\Tilde{\psi}}$ where $\hat{U}_0(t)$ satisfies (let $\hbar = 1$)
\begin{equation}
    \pdv{\hat{U}_0(t)}{t} = -i\hat{H}_0(t) \hat{U}_0(t).
\end{equation}
Then, the Schr\"{o}dinger equation in the interaction picture is
\begin{equation}
    i\pdv{\ket*{\Tilde{\psi}}}{t} = \hat{U}_0(t)^\dagger \delta \hat{H}(t) \hat{U}_0(t) \ket*{\Tilde{\psi}(t)} = h(t) \hat{O}_H(t) \ket*{\Tilde{\psi}}
\end{equation}
where $\hat{O}_H(t)$ is the operator $\hat{O}(t)$ in the interaction picture, $\hat{O}_H(t) = \hat{U}_0^\dagger(t) \hat{O}(t) \hat{U}_0(t)$.

Next, we perform a first-order time-dependent perturbation calculation. We replace $\ket*{\Tilde{\psi}}$ with $\ket{\psi}$ and $\hat{O}_H$ with $\hat{O}$ for simplicity in the following. Let $\ket{\psi(t)} = \ket{\psi_0} + \ket{\psi_1 (t)}$ where $\ket{\psi_0} = \ket{\psi(0)}$ is the initial state and $\ket{\psi_1}$ is a perturbation. Then by the first order perturbation calculation,
\begin{equation}
    \ket{\psi_1(t)} = -i\int_0^t d\tau \ h(\tau) \hat{O}(\tau) \ket{\psi_0}.
\end{equation}

Assuming the noise amplitude $h(t)$ is described by an one-sided PSD $S(f)$, the time trace of $h(t)$ can be expressed (with abuse of notation $df$) as
\begin{equation}
    h(t) = \sum_f \sqrt{2S(f)df} \cos(2\pi f t + \phi_f),
\end{equation}
where $\phi_f$ is a random phase for each frequency component $f$. The phases for any two different frequencies are independent. We have
\begin{equation}
    \begin{split}
    &\ket{\psi_1(t)} \\= &-\sqrt{2} i \int_0^t d\tau \sum_f \sqrt{S(f)df}\cos(2\pi f \tau + \phi_f)\hat{O}(\tau) \ket{\psi_0}.
    \end{split}
\end{equation}

The fidelity can be expanded to the leading order as 
\begin{equation}
    F = \dfrac{|\braket{\psi_0}{\psi}|^2}{\braket{\psi}{\psi}} = 1 - (\braket{\psi_1}{\psi_1} - |\braket{\psi_0}{\psi_1}|^2) + \cdots.
\end{equation}
We need to evaluate the two quadratic terms of $\ket{\psi_1}$. This is the autocorrelation function of the noise operator, which intuitively gives a term linear in $S(f)$:
\begin{widetext}
    \begin{equation}
    \braket{\psi_1}{\psi_1} = 2\iint dt d\tau \sum_{f,g} \sqrt{S(f)S(g)dfdg} \langle \cos(2\pi f t + \phi_f)\cos(2\pi g \tau + \phi_g) \rangle_{\phi_f, \phi_g} \bra{\psi_0} \hat{O}(\tau)\hat{O}(t) \ket{\psi_0},
\end{equation}
\begin{equation}
    |\braket{\psi_0}{\psi_1}|^2 = 2\iint dt d\tau \sum_{f,g} \sqrt{S(f)S(g)dfdg} \langle \cos(2\pi f t + \phi_f)\cos(2\pi g \tau + \phi_g) \rangle_{\phi_f, \phi_g} \bra{\psi_0} \hat{O}(\tau) \ket{\psi_0} \bra{\psi_0} \hat{O}(t) \ket{\psi_0}.
\end{equation}
\end{widetext}

Both equations involve averaging over $\phi_f\ \text{and}\ \phi_g$. Since $\phi_f$ for different frequencies are mutually independent, the averaged product of two cosine functions vanishes unless $f=g$:

\begin{equation}
\begin{split}
    &\langle \cos(2\pi f t + \phi_f)\cos(2\pi g \tau + \phi_g) \rangle_{\phi_f, \phi_g} \\= &\dfrac{1}{2} \cos(2\pi f(t-\tau)) \delta_{f,g}.
\end{split}
\end{equation}

\begin{widetext}
Hence,
    \begin{equation}
    F = 1-\int_0^T\int_0^T dt d\tau \int_{0}^\infty df S(f) \cos(2\pi f(t-\tau)) [\langle \hat{O}(t)\hat{O}(\tau) \rangle - \langle \hat{O}(t) \rangle\langle \hat{O}(\tau) \rangle]
    \end{equation}
\end{widetext}
where $\langle \hat{O} \rangle$ is the shorthand notation for $\bra{\psi_0} \hat{O} \ket{\psi_0}$, and $T$ is the total evolution time.

The infidelity is, to the first order, linear in $S(f)$. Notice that this can be understood as integrating $S(f)$ weighted by a filter function $I(f)$, which is the sensitivity of fidelity to a delta function PSD at frequency $f$, as explained in the main text. Explicitly, the response function is
\begin{widetext}
    \begin{equation}\label{eq:response_single_state}
    I(f) = \int_0^T dt \int_0^T d\tau \cos(2\pi f(t-\tau)) [\langle \hat{O}(t)\hat{O}(\tau) \rangle - \langle \hat{O}(t) \rangle\langle \hat{O}(\tau) \rangle].
    \end{equation}
\end{widetext}

The formula above evaluates the quantum process fidelity given the initial state $\ket{\psi_0}$. We also compute the average gate fidelity over a set of states. We distinguish two sets of scenarios: averaging over a finite set of states or averaging over an input Hilbert space.

To compute the average response function over a finite set of input states, we compute the mean of response functions for all states. 

To compute the response function averaged over Haar random measure of an input Hilbert space, we evaluate the expectation value of the above formula by integrating over the Haar measure on the $D$-dimensional Hilbert space (\textit{e.g.,} $D=4$ for two-qubit Haar random states and $D=3$ for two-qubit symmetric Haar random states)
\begin{widetext}
\begin{equation}\label{eq:response_avg}
\begin{split}
    I_\mathrm{avg}(f) = \int_0^T dt \int_0^T d\tau &\cos (2\pi f(t-\tau))
    \left\{\dfrac{1}{D} \mathrm{Tr}\left[\hat{O}(t)\hat{O}(\tau)P\right] - \dfrac{1}{D(D+1)} \left[\mathrm{Tr}\left[\hat{O}(t)P\hat{O}(\tau)P\right] + \mathrm{Tr}\left[\hat{O}(t)P\right]\mathrm{Tr}\left[\hat{O}(\tau)P\right]\right]\right\}
\end{split}
\end{equation}
\end{widetext}
where $P$ is the projector from the full Hilbert space to the input Hilbert space.

\section{Analytic prediction of Rydberg spin-lock experiment}\label{Appendix:spin_lock}
\renewcommand{\theequation}{H\arabic{equation}}
We give further details on the spin-lock experiment and how it is used to probe frequency noise PSD using an atomic signal.

For simplicity, we first assume that the preparation and readout pulses are noise-free, and there is no spontaneous Rydberg decay. In the Rydberg manifold (the \clock\ state and the Rydberg state in Fig.~\ref{fig:overview}d), we initialize all atoms in an eigenstate of $\hat{X}$ and turn on a continuous \textit{locking} drive along the $\hat{X}$-axis for a variable time. Then we apply a $\pi/2$ pulse about the $\hat{Y}$-axis, which transfers all atoms into the Rydberg state in the absence of Rydberg laser frequency noise. In the presence of noise, the probability of returning to the Rydberg state decays over time. Finally, we perform a push-out on the atoms in the Rydberg state with an auto-ionization pulse~\cite{Madjarov2020} and image the atoms remaining in the \clock\ state to obtain this atomic signal, characterized by a decay rate which is predominantly sensitive to frequency noise $S_\nu(f)$ around this locking Rabi frequency, and a converging value of $1/2$. 

For a decay-free experiment, we compute the frequency response with Eq.~\ref{eq:response_func} by plugging in $\ket{\psi_0} = \ket{+}$, $\hat{O} = -\pi \hat{\sigma}_z$, and $\hat{H}_0 = \Omega \hat{\sigma}_x/2$. The response to laser frequency noise is:
\begin{equation}
\begin{split}
      I_\nu(f) = \dfrac{1}{2}\pi^2 t^2 &\left\{\mathrm{sinc}^2\left[\left(\frac{\Omega}{2} + \pi f\right) t\right] \right.\\
      + &\left.\mathrm{sinc}^2\left[\left(\frac{\Omega}{2} - \pi f\right) t\right] \right\}.  
\end{split}
\end{equation}
where $t$ is the probe time.

In the case that $S_{\nu}$ is upper bounded, in the long time limit, $\lim_{t\rightarrow\infty} \frac{1}{\pi t} \frac{\sin^2[(x-a)t]}{(x-a)^2}=\delta(x-a)$. Then, the response function simplifies to
\begin{equation}
    I_\nu (f) \approx \dfrac{1}{2} \pi^2 \delta\left(f-\dfrac{\Omega}{2\pi}\right) t
\end{equation}
Hence, if we choose a locking Rabi frequency $\Omega$ and fit the spin-lock signal to a trial function $(1+\mathrm{e}^{-\Gamma t})/2$, then $\Gamma = \pi^2 S_{\nu}(\Omega/(2\pi))$.

The Rydberg decay adds an additional decay to the spin-lock signal, which corresponds to the decay rate when the state is on the equator, $1/(2\tau)$, where $\tau$ is the Rydberg state lifetime. In our experimental setup, another problem comes with this decay: atoms in the Rydberg state can decay into states that contribute falsely to the final imaging, as seen in Appendix~\ref{Appendix:BrightState}. This adds another layer of difficulty to the measured signal, as we typically assume that the Rydberg population is $1-P_{\text{imaged}}$, and now $P_{\text{imaged}}$ contains a part of the population decayed from the Rydberg state. To circumvent this issue, we perform another separate spin-lock experiment with the difference of applying a $\pi/2$ pulse about the $-\hat{Y}$-axis, instead of $\hat{Y}$-axis. In the absence of errors, atoms are transferred back to the \clock\ state. By taking the difference of the signals from the two different sequences, we get rid of this false contributions to the final image. In the experiment, we fit this difference signal to a trial function $a_0 + b_0\times\mathrm{e}^{-\Gamma t}$ and quote $\Gamma$ as the spin-lock decay rate.

As discussed in the main text, at various locking Rabi frequencies, we compare the decay rate from a full error model simulation and the analytical prediction from FRT (red markers and dark green line in Fig.~\ref{fig:spin_lock}). The agreement confirms that the spin-lock experiment indeed provides a measurement of the PSD at a certain frequency of interest.

We further compare these results to the experimental results (purple markers in Fig.~\ref{fig:spin_lock}), obtained from the procedure explained above. From the consistency, we probe the strength of laser frequency noise directly with an atomic signal at several test frequencies.

\section{Experimentally learning the response function}\label{Appendix:exp_response_function}

\begin{figure}[ht!]
    \centering
    \includegraphics[width=\columnwidth]{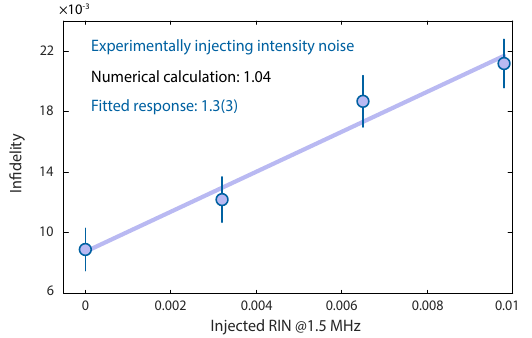}
    \caption{\textbf{Experimentally learning the response function.} The response function can also be studied experimentally. As an illustrative example, we experimentally extract the value in the time-optimal CZ gate fidelity response function of the laser intensity noise. We inject intensity modulations at a fixed frequency ($1.5$~MHz) during the time-optimal CZ gates, performed at Rabi frequency of 3~MHz. By fitting the gate infidelity linearly to the injected noise strength, we extract the value at this frequency $I_{\text{exp}}(1.5~\text{MHz}) = 1.3(3)$, in agreement with the numerical calculation $I_{\text{num}}(1.5~\text{MHz})=1.04$ (Fig.~\ref{fig:linear_response}a, lower).} 
    \vspace{-0.3cm}
    \label{SI_exp_response_int}
\end{figure}

By definition, the response $I(f)$ at a certain frequency $f$ is the ratio of the induced infidelity to the delta function PSD strength at frequency $f$. Hence, other than numerically calculating the response function, which can be demanding when the system size is large, we can directly measure $I(f)$ from experiment at certain $f$. 

We show a proof-of-principle of such measurement. We measure the intensity response at $f=1.5\ \mathrm{MHz}$ of the time-optimal gate with Rabi frequency $3.0\ \mathrm{MHz}$ (Fig.~\ref{SI_exp_response_int}). We inject a monochromatic intensity modulation at 1.5~MHz with the pulsing AOM with random initial phase relative to the pulse and measure the gate infidelity. Then we fit the infidelity versus modulation strength to a linear function and extract the slope, which is the linear coefficient of the intensity response at the modulation frequency. The measurement is consistent with the numerical prediction from the closed-form evaluation (Eq.~\ref{eq:response_func}). 

\section{Response functions for realistic gates}\label{Appendix:realistic_gate}
\renewcommand{\theequation}{J\arabic{equation}}
FRT is widely applicable, not limited to the simplified gate dynamics as described in Eq.~\ref{eq:gateHamiltonian}, but also holds for a \textit{realistic} gate with finite rise time, self-light shift, and finite blockade interaction. We note that all numerical results from FRT presented in this work are based on this simplified \textit{ideal} gate dynamics. The Hamiltonian that we simulate for a \textit{realistic} gate (used in the full \textit{ab initio} error model simulation) is
\begin{equation}\label{eq:gate_Ham_realistic}
\begin{split}
    \hat{H} &= \dfrac{\Omega(t)}{2} \sum_{i=1}^2 (e^{-i\varphi(t)}\ketbra{1_i}{r_i}+h.c.)\\
    &- \left(\Delta  - \kappa_r \Omega(t)^2\right) \sum_{i=1}^2 \ketbra{r_i} \\
    &+ \kappa_g \Omega(t)^2 \sum_{i=1}^2 \ketbra{0_i} + B \ketbra{r_1r_2},
\end{split}
\end{equation}
where $\Omega(t)$ is a smooth function with finite rise and fall time, limited by AOM (as depicted in Fig.~\ref{fig:overview}e, as opposed to the constant $\Omega$ in Eq.~\ref{eq:gateHamiltonian}), which are measured in experiment with a Si-amplified photodiode; $\Delta$ is a fixed detuning, as part of the time-optimal CZ gate parameter set; $\kappa_r$ and $\kappa_g$ are the differential polarizabilities of the Rydberg state and $^1\text{S}_0$, respectively, relative to the \clock\ state, which are independently measured. Symbols that are not explicitly defined here represent the same as in Eq.~\ref{eq:gateHamiltonian}.

We then identify the noise operators for frequency noise and intensity noise:
\begin{align}
    \hat{O}_\nu (t) &= -2\pi \sum_{i=1}^2 \ketbra{r_i},\\
    \hat{O}_I(t) &= \dfrac{\Omega(t)}{4} \sum_{i=1}^2 (e^{-i\varphi(t)}\ketbra{1_i}{r_i}+h.c.) \notag\\ 
    &+ \Omega(t)^2 \sum_{i=1}^2 (\kappa_g \ketbra{0_i} + \kappa_r \ketbra{r_i}).
\end{align}

Then, the calculation of corresponding response functions follows from Eq.~\ref{eq:response_avg}. Qualitatively, the gate time is longer due to the rise/fall time, leading to higher infidelity than the \textit{ideal} gates used in the scaling analysis which assumes zero rise/fall time (Fig.~\ref{fig:overview}f and Fig.~\ref{fig:error_scaling}).

\section{Generalization to two-photon transition}\label{Appendix:two_photon}
\renewcommand{\theequation}{K\arabic{equation}}
We briefly describe a generalization to CZ gates mediated by a two-photon Rydberg transition. We model the two-photon Rydberg transition with a four-level system per atom, containing the qubit states ($\ket{0}, \ket{1}$), the Rydberg state $\ket{r}$, and the intermediate state $\ket{e}$.
We consider the following Hamiltonian for CZ gates implemented with a two-photon transition
\begin{equation}
\begin{split}
    \hat{H} &= \dfrac{\Omega_1(t)}{2} \sum_{i=1}^2 (\ketbra{1_i}{e_i}+h.c.) + \dfrac{\Omega_2(t)}{2} \sum_{i=1}^2 (\ketbra{e_i}{r_i}+h.c.)\\
    &+ (\kappa_{0,1}\Omega_1(t)^2 + \kappa_{0,2}\Omega_2(t)^2) \sum_{i=1}^{2} \ketbra{0_i}\\
    &+ (\kappa_{1,1}\Omega_1(t)^2 + \kappa_{1,2}\Omega_2(t)^2) \sum_{i=1}^{2} \ketbra{1_i} \\
    &- (\Delta + \delta_1(t))\sum_{i=1}^2 \ketbra{e_i} \\
    &- (\delta_1(t) + \delta_2(t) - \kappa_{r,1}\Omega_1(t)^2 - \kappa_{r,2}\Omega_2(t)^2) \sum_{i=1}^2 \ketbra{r_i}\\
    &+ B \ketbra{r_1r_2},
\end{split}
\end{equation}
where $i=1,2$ is the atom label, $\Omega_1(t)$ and $\Omega_2(t)$ are the Rabi frequencies of each arm, $\Delta$ is the intermediate state detuning, $\delta_1(t)$ and $\delta_2(t)$ are the additional detuning modulation of each arm (between $\ket{1}, \ket{e}$ and between $\ket{e}, \ket{r}$, respectively), $\kappa_{0,j}, \kappa_{1,j}, \kappa_{r,j}$ represent relative polarizabilities of state $\ket{0}, \ket{1}, \ket{r}$ with respect to the intermediate state $\ket{e}$ due to laser $j=1, 2$, and $B$ is the Rydberg blockade interaction. 

The response to laser intensity noise and laser frequency noise of both lasers, assuming these noise sources are mutually independent, is calculated with Eq.~\ref{eq:response_avg} by identifying the corresponding noise operators $\hat{O}$. The intensity noise operators for arm 1 and 2 are
\begin{align}\label{eq:intensity_4level}
    \hat{O}_{I1}(t) &= \dfrac{\Omega_1(t)}{4} \sum_{i=1}^2 (\ketbra{1_i}{e_i} + h.c.) \notag\\
    & + \Omega_1(t)^2 \sum_{i=1}^2 (\kappa_{0,1} \ketbra{0_i} + \kappa_{1,1} \ketbra{1_i} + \kappa_{r,1} \ketbra{r_i}),\\
    \hat{O}_{I2}(t) &= \dfrac{\Omega_2(t)}{4} \sum_{i=1}^2 (\ketbra{e_i}{r_i} + h.c.) \notag\\
    & + \Omega_2(t)^2 \sum_{i=1}^2 (\kappa_{0,2} \ketbra{0_i} + \kappa_{1,2} \ketbra{1_i} + \kappa_{r,2} \ketbra{r_i}),
\end{align}
and the frequency noise operators for arm 1 and 2 are
\begin{align}\label{eq:frequency_4level}
    \hat{O}_{\nu 1}(t) &= -2\pi \sum_{i=1}^2 (\ketbra{e_i} + \ketbra{r_i}),\\
    \hat{O}_{\nu 2}(t) &= -2\pi \sum_{i=1}^2 \ketbra{r_i}. \label{eq:frequency_4level2}
\end{align}
By plugging these operators back into Eq.~\ref{eq:response_avg}, one can compute all four response functions. It follows that the infidelity of the gates, to the leading order, is
\smallskip
\begin{equation}
\begin{split}
    1-F &= \int_0^\infty df S_{\nu 1}(f) I_{\nu 1}(f) + \int_0^\infty df S_{\nu 2}(f) I_{\nu 2}(f)\\
    &+\int_0^\infty df S_{I 1}(f) I_{I 1}(f) + \int_0^\infty df S_{I 2}(f) I_{I 2}(f)
\end{split}
\end{equation}
where $S_{\nu 1}$ and $S_{\nu 2}$ are frequency PSDs of the two lasers, $S_{I 1}$ and $S_{I 2}$ are the RIN PSDs of the two lasers, and $I$ are the corresponding response functions.

Recently, such two-photon Rydberg CZ gates~\cite{Graham2019,Levine2019,Evered2023} have been demonstrated in the limit of $\Delta \gg B, \Omega_1, \Omega_2 \gg \delta_1, \delta_2 \sim \frac{\Omega_1 \Omega_2}{2\Delta}$ and the intermediate state detuning greater than all light shifts. Under this limit, there exists an effective Rabi frequency of $\frac{\Omega_1 \Omega_2}{2\Delta}$ and a self-light shift of $\frac{\Omega_1^2 - \Omega_2^2}{4\Delta}$ between $\ket{1}$ and $\ket{r}$, and the intermediate state is adiabatically eliminated. On top of that, we further let $\Tilde{\kappa}_{0,j} = \kappa_{0,j} - \kappa_{1,j}$ and $\Tilde{\kappa}_{r,j} = \kappa_{r,j} - \kappa_{1,j}$ denote the polarizabilities of $\ket{0}$ and $\ket{r}$ relative to $\ket{1}$. The effective Hamiltonian is

\begin{equation}
\begin{split}
    \hat{H}_\mathrm{eff} &= \dfrac{\Omega_1(t)\Omega_2(t)}{4\Delta} \sum_{i=1}^2 (\ketbra{1_i}{r_i}+h.c.) \\
    &+ \Omega_1(t)^2 \sum_{i=1}^2 \left(\Tilde{\kappa}_{0,1} \ketbra{0_i} + \Tilde{\kappa}_{r,1} \ketbra{r_i} \right) \\
    &+ \Omega_2(t)^2 \sum_{i=1}^2 \left(\Tilde{\kappa}_{0,2} \ketbra{0_i} + \Tilde{\kappa}_{r,2} \ketbra{r_i} \right) \\
    &- \left[\delta_1(t)+\delta_2(t)+ \dfrac{\Omega_1(t)^2 - \Omega_2(t)^2}{4\Delta} \right] \sum_{i=1}^2 \ketbra{r_i} \\
    &+ B \ketbra{r_1r_2}
\end{split}
\end{equation}

Intuitively, following the treatment of one-photon transition and viewing these noise sources directly as perturbations on the effective Rabi frequency, one could write down the intensity noise operators and frequency noise operators as follow:

\begin{widetext}
\begin{align}\label{eq:operator_3level}
    \vspace{-0.2cm}
    \hat{O}_{I1}(t) &= \dfrac{\Omega_1(t)\Omega_2(t)}{8\Delta} \sum_{i=1}^2 (\ketbra{1_i}{r_i} + h.c.) + \Tilde{\kappa}_{0,1} \Omega_1(t)^2 \sum_{i=1}^2 \ketbra{0_i} +\left(\Tilde{\kappa}_{r, 1} - \dfrac{1}{4\Delta} \right) \Omega_1(t)^2  \sum_{i=1}^2 \ketbra{r_i},\\
    \hat{O}_{I2}(t) &= \dfrac{\Omega_1(t)\Omega_2(t)}{8\Delta} \sum_{i=1}^2 (\ketbra{1_i}{r_i} + h.c.) + \Tilde{\kappa}_{0,2} \Omega_2(t)^2 \sum_{i=1}^2 \ketbra{0_i} +\left(\Tilde{\kappa}_{r,2} + \dfrac{1}{4\Delta} \right) \Omega_2(t)^2  \sum_{i=1}^2 \ketbra{r_i},\\
    \hat{O}_{\nu 1}(t) &= -2\pi \sum_{i=1}^2 \ketbra{r_i},\\
    \hat{O}_{\nu 2}(t) &= -2\pi \sum_{i=1}^2 \ketbra{r_i}. \label{eq:operator_3level_final}
\end{align}
\end{widetext}

This reduces the evaluation of response functions from a 4-level to a 3-level system. Hence, the analysis is then similar to that of a single-photon Rydberg transition, except for the existence of these four (instead of two) noise operators. The corresponding noise PSDs for these four operators are RIN PSD of laser 1, RIN PSD of laser 2, frequency noise PSD of laser 1, and frequency noise PSD of laser 2, respectively.

\section{Approximate forms of universal response functions}\label{Appendix:ApproxForms}
We give approximate forms (6-parameter fits) of the universal response functions for \textit{ideal} time-optimal gates with both fidelity metrics $F_\mathrm{Haar}$ and $F_\mathrm{Sym}$ (Fig.~\ref{SI_response_compare}).

Frequency response for $F_\mathrm{Haar}$:
\begin{align*}
    \dfrac{g_\nu (x)}{(2\pi)^2} = a\,\exp \left[-\left(\frac{x-b}{c}\right)^2\right] + d\,\exp \left[-\left(\frac{x-e}{f}\right)^2\right],\\
    a = 2.910, b = -0.02715, c = 0.5874, \\
    d = 3.022, e = 1.179, f = 0.5337.
\end{align*}

Intensity response for $F_\mathrm{Haar}$:
\begin{align*}
    g_I (x) = a\dfrac{1 + d\, \tanh{[e(x-f)]}}{1+\exp[b(x-c)]},\\
    a = 1.187, b = 6.423, c = 0.7670, \\
    d = 0.07678, e = 5.528, f = 0.2381.
\end{align*}

Frequency response for $F_\mathrm{Sym}$:
\begin{align*}
    \dfrac{g_\nu (x)}{(2\pi)^2} = a\,\exp \left[-\left(\frac{x-b}{c}\right)^2\right] + d\,\exp \left[-\left(\frac{x-e}{f}\right)^2\right],\\
    a = 3.062, b = -0.01507, c = 0.5588,  \\
    d = 2.843, e = 1.232, f = 0.5339.
\end{align*}

Intensity response for $F_\mathrm{Sym}$:
\begin{align*}
    g_I (x) = a\dfrac{1 + d\, \tanh{[e(x-f)]}}{1+\exp[b(x-c)]},\\
    a = 1.218, b = 5.790, c = 0.7580, \\
    d = 0.03630, e = 5.647, f = 0.2054.
\end{align*}

\input{main.bbl}

\end{document}

%% file: main.bbl
%